\documentclass[sigconf,final]{acmart}
\usepackage{amsmath,amsfonts}
\usepackage{esvect}
\usepackage{graphicx}
\usepackage{subfigure}
\usepackage{booktabs}
\usepackage{tabularx}
\usepackage{multirow}
\usepackage{balance}
\usepackage{xcolor}
\usepackage{enumitem}
\usepackage{mathtools}
\usepackage{wrapfig}
\usepackage{ragged2e}
\usepackage[all]{nowidow}
\usepackage{stfloats}
\usepackage{anyfontsize}
\usepackage{color}
\usepackage{colortbl}
\usepackage{mathtools}
\usepackage[linesnumbered,ruled,vlined]{algorithm2e}
\usepackage{syntax}

\makeatletter %only in the preamble
\DeclareRobustCommand\newttfamily
{\not@math@alphabet\ttfamily\mathtt
  \fontfamily\ttdefault\selectfont\hyphenchar\font=`\-\relax}
\makeatother %only in the preamble
\DeclareTextFontCommand{\texttth}{\newttfamily}
\usepackage{listings}

\definecolor{javared}{rgb}{0.6,0,0} % for strings
\definecolor{javagreen}{rgb}{0.25,0.5,0.35} % comments
\definecolor{javapurple}{rgb}{0.5,0,0.35} % keywords
\definecolor{javadocblue}{rgb}{0.25,0.35,0.75} % javadoc
\definecolor{javagrey}{rgb}{0.46,0.45,0.48} % annotations

\newcommand\resq[1]{
\noindent 
\fcolorbox{green!40!black}{green!5}{\noindent 
 \parbox{0.98\columnwidth}{\noindent  #1}}\\
}
\lstdefinestyle{Spec}{
	language=Java, % choose the language of the code
	keywordstyle=\color{javapurple}, %\bfseries,
	stringstyle=\color{javared},
	commentstyle=\color{javagreen},
	morecomment=[s][\color{javadocblue}]{/**}{*/},
	morecomment=[l][\color{javagrey}]{@},
	morecomment=[l][\color{javagrey}]{//@},
  morecomment=[l][\color{javagrey}]{/*@},
  morecomment=[l][\color{javagrey}]{*/},
	basicstyle=\ttfamily\footnotesize,
	breaklines=true,
	tabsize=2,
	frame=single,
	mathescape,
	numbers=left,
	xleftmargin=2.5em,
	xrightmargin=0.5em,
	frame=single,
	framexleftmargin=2em,
	morekeywords={,duration,simulationNodes,ms,Platform,CPU,memory,SimulationNode,platform,cloud,Cloud,IP,port,protocol,b,Simulator,username,password,quanta,step,Device,period,payload,speed,devices,locationIP, G, EdgeDevices,workload, inToOut, EdgeDevice,type, }
	escapeinside={\%*}{*)} % if you want to add a comment within your code
    columns=flexible,
	escapechar=?,
}

\lstdefinestyle{myCustomMatlabStyle}{
  language=Matlab,
  numbers=left,
  stepnumber=1,
  numbersep=10pt,
  tabsize=4,
  showspaces=false,
  showstringspaces=false
}

\usepackage{framed}
\definecolor{light-gray}{gray}{0.9}
\usepackage[framemethod=TikZ]{mdframed}
\mdfsetup{skipabove=5pt,skipbelow=3pt}
\usepackage{lipsum}
\mdfdefinestyle{MyFrame}{%
	linecolor=black,
	outerlinewidth=0.15pt,
	roundcorner=3pt,
	innertopmargin=2pt,
	innerbottommargin=2pt,
	innerrightmargin=4pt,
	innerleftmargin=4pt,
	backgroundcolor=light-gray}
\newcolumntype{K}[1]{>{\centering\arraybackslash}p{#1}}
\copyrightyear{2022}
\acmYear{2022}
\setcopyright{acmcopyright}\acmConference[MODELS '22]{ACM/IEEE 25th International Conference on Model Driven Engineering Languages and Systems}{October 23--28, 2022}{Montreal, QC, Canada}
\acmBooktitle{ACM/IEEE 25th International Conference on Model Driven Engineering Languages and Systems (MODELS '22), October 23--28, 2022, Montreal, QC, Canada}
\acmPrice{15.00}
\acmDOI{10.1145/3550355.3552405}
\acmISBN{978-1-4503-9466-6/22/10}

\newcommand\lang{\textsf{IoTECS}}

\begin{document}

%\title{SIMTEC: An IoT Edge to Cloud Simulation Language for Model Testing of IoT Products}

\title[]{A Domain-Specific Language for Simulation-Based Testing of \\ IoT Edge-to-Cloud Solutions}

\author{Jia Li, Shiva Nejati, Mehrdad Sabetzadeh}
\affiliation{
  \institution{University of Ottawa}
  \country{Canada}
}
\email{{jli714, snejati,  m.sabetzadeh}@uottawa.ca}

%\author{}
%\affiliation{%
%  \institution{University of Ottawa}
%  \country{Canada}
%}
%\email{snejati@uottawa.ca}

%\author{}
%\affiliation{%
%\institution{University of Ottawa}
%  \country{Canada}
%}
%\email{}

\author{Michael McCallen}
\affiliation{
  \institution{Cheetah Networks}
  \country{Canada} 
}
\email{mccallen@cheetahnetworks.com}

\begin{abstract}
The Internet of things (IoT) is increasingly prevalent in domains such as emergency response, smart cities and autonomous vehicles. Simulation plays a key role in the testing of IoT systems, noting that field testing of a complete IoT product may be infeasible or prohibitively expensive. In this paper, we propose a \emph{domain-specific language (DSL)} for generating edge-to-cloud simulators. An edge-to-cloud simulator executes the functionality of a large array of edge devices that communicate with cloud applications. Our DSL, named \lang, is the result of a collaborative project with an IoT analytics company, Cheetah Networks. 
The industrial use case that motivates \lang\ is ensuring the scalability of cloud applications by putting them under extreme loads from IoT devices connected to the edge. We implement \lang\ using Xtext and empirically evaluate its usefulness. We further reflect on the lessons learned.\vspace*{-.5em}
\end{abstract}

\maketitle

\vspace*{-.25em}
\section{Introduction}
\label{sec:intro}
%simulation is important 
%Simulation is an important method for the validation and verification of complex software-intensive systems. %Simulation-based solutions play a key role in many recent technological and engineering advancements.
Simulation is an important validation and verification activity during the requirements and design stages, and before a proposed system has been built. For certain types of systems, simulation also plays a key role in later stages of development, e.g., when one needs to exercise scenarios that are too risky or too expensive to run in real-world deployments.
%Simulation is used in a wide range of analytical tasks, including decision making, safety and security assurance, testing, and reliability assessment~\cite{??}.
Notably, simulation is the technique of choice for the validation and verification of systems that have high levels of autonomy, e.g., autonomous vehicles~\cite{Al-Sultan:14,Borg:21,Ahlgren:21}, or massive connectivity, e.g., the Internet of things (IoT)~\cite{Shin:20,Li:22}. 

%we build an IoT edge to cloud simulator. What is IoT?
%Simulators are an important enabler for model-in-the-loop testing of complex systems.

This paper is concerned with simulation in the context of IoT. IoT envisions complex systems that interconnect large numbers of smart devices, embedded with sensors and actuators, through the Internet or other network technology~\cite{IoT17}. IoT is already pervasive in several application domains, e.g., emergency response, smart cities, smart agriculture and autonomous vehicles, to name a few. 
%
%
%In this paper, we concern ourselves with a particular facet of simulation in the context of IoT. We aim to extend the use of simualtors for model in the loop testing to the domain of IoT.
%our focus is on edge to cloud agent simulation
%
IoT simulation has numerous facets, including simulation of IoT sensors and actuators~\cite{Kertesz:19,AmazonIoT,Zeng:17}, simulation of IoT edge devices~\cite{Jha:20,Sonmez:17} and simulation of IoT networks~\cite{Kliazovich:12,Osterlind:06,boulis:11,Mininet}. 
Our work in this paper focuses on simulating \emph{edge devices and their interactions with cloud applications}. An edge device is a generic term referring to any device that serves as an entry point to a network. In IoT, edge devices typically run applications and protocols for managing groups of sensors and actuators, performing edge computing~\cite{Elazhary:19}, and sending and receiving data to and from cloud applications.
%A cloud application provides cloud-computing services based on the data received from the edge devices. 

Specifically, we propose in this paper a model-based approach for creating \emph{edge-to-cloud simulators}. An edge-to-cloud simulator executes the functionality of a large number of edge devices that run in parallel and communicate with cloud applications either synchronously or asynchronously~\cite{Elazhary:19}. Edge-to-cloud simulators may be employed for purposes such as testing and decision making, e.g., about how frequently edge devices should communicate with the cloud, and how to schedule the data flow from IoT devices so that the overall network traffic remains stable and burst-free~\cite{Wang:21}.

%
%Specifically, we develop a versatile and effective approach for simulating the communication between edge devices and cloud applications. 
Our work in this paper addresses a specific yet important need in IoT: \emph{Scale testing of cloud applications}. Scale testing is concerned with understanding how a system behaves under an extreme workload and at the upper limits of its capacity~\cite{Chan:04}. In our context, the workload originates from the edge devices and is destined for processing and storage by the cloud. Existing feature-rich IoT simulators, e.g.~\cite{Jha:20,Sonmez:17}, are not good fits for cloud scale testing, noting that a feature-rich simulator may require substantial computational resources (CPU, memory and storage) to simulate thousands of devices. Many IoT providers use hosted services for simulation. The more resource-intensive the simulator, the more difficult it is to \hbox{set up} and the higher is the cost of running it for scale testing.

\begin{figure}[!t]
    \includegraphics[width=.65\linewidth]{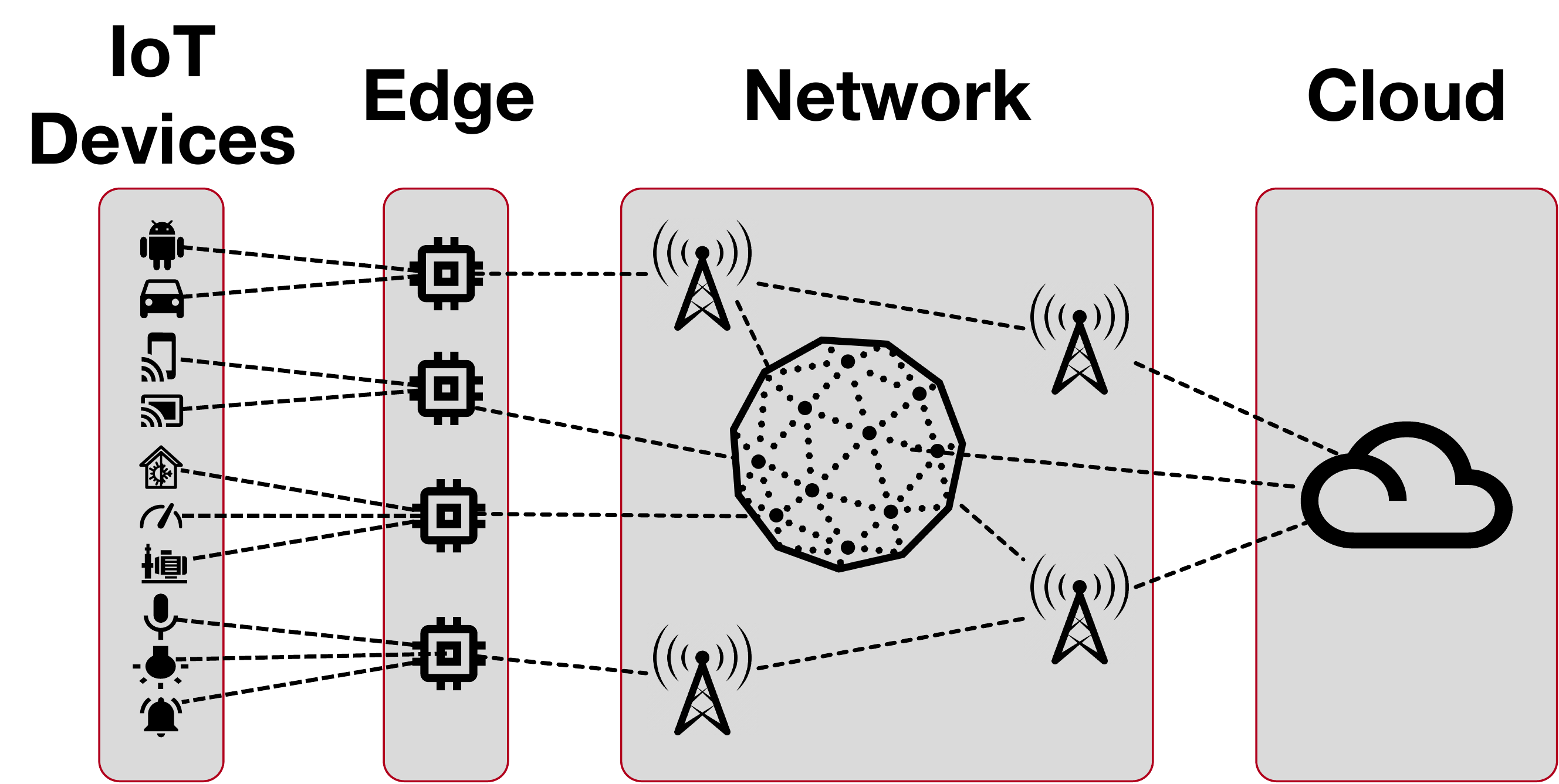}
    \vspace*{-.5em}
    \caption{Conceptual view of a typical IoT System.}\label{fig:edge-cloud}
    \vspace*{-1.8em}
\end{figure}

%Why we focus on edge to cloud agent?
Figure~\ref{fig:edge-cloud} shows a schematic view of a typical IoT system. In a real IoT system, a network sits in between the edge and cloud layers. In our work, we need to ensure that the network does not act as a limiter for how far simulated edge devices can stress the cloud. We therefore connect the simulated edge layer to the cloud via an ideal network with no loss and minimally low transmission time. For experimentation purposes, we create this ideal network using a single %high-throughput 
network switch, discussed in Section~\ref{subsec:expdesign}. 
%that sit in between the edge and cloud layers, and instead focus on simulating the behaviour of edge devices and cloud applications layers and their communication. 

\textbf{Contribution.} Building a well-functioning edge-to-cloud simulator involves various technical considerations, including proper use of communication protocols, managing the massive concurrency of simulated edge devices, and efficient use of computing resources. Due to these considerations, building an edge-to-cloud simulator from scratch can be a daunting task. To facilitate the construction of edge-to-cloud simulators, we propose a \emph{domain-specific language (DSL)}, named \lang\ (the \emph{IoT Edge-to-Cloud Simulation} language). \lang\ provides abstractions that shield software engineers from several complexities of edge-to-cloud simulation, while still affording good customizability for different systems and different resource constraints. Specifications (models) written in \lang\ are automatically translatable into operational Java-based simulators. \lang\ has been developed in response to industrial needs and is the outcome of a collaborative research activity with Cheetah Networks -- a leading provider of IoT analytics solutions. \lang\ has been already adopted by Cheetah Networks for stress testing some of their cloud applications.

Following ideas and guidelines from the literature on multi-agent simulation~\cite{Scheffer:95}, \lang\ employs a hierarchical architecture for grouping large arrays of simulated edge devices. The resulting groups can subsequently be emulated through containerization, virtualization, or a combination thereof. \lang\ further supports configurable time gaps for orchestrating the execution of parallel edge devices; this is to ensure that messages between the edge devices and the cloud applications are communicated at a rate that is in line with real-world expectations in a specific IoT system.

\textbf{Evaluation.} We have applied \lang\ for stress testing a benchmark cloud as well as industrial cloud applications at our partner, Cheetah Networks. To evaluate \lang, we first ensure that it can reliably simulate a large number of parallel edge devices without incurring data loss. We scale simulation to generate 24000 packets per second using a single conventional laptop computer. These packets represent the data generated by 12000 (simulated) IoT sensors and actuators that communicate with the cloud every 500 milliseconds. We show how \lang\ can help engineers determine in a systematic way the number of IoT and edge devices that the cloud applications under test can handle. We further reflect on our experience and lessons learned using \lang\ in an industrial setting.

\iffalse
setting and report on the adoption of \lang\ and its successful use within our partner company. 
\fi

\iffalse
%applicatoins of our IoT simulator. 
Our DSL language translates into an operational simulator that can be used for a number of purposes: (1)~demonstrating the scalability of cloud servers and their ability to respond and send messages (to test scaling of cloud agents),
(2)~demonstrating the scalability of an entire IoT applications in terms of the number of devices and edge agents that can be supported, and (3)~enabling engineers to make decisions about the design of their applications (e.g., the maximum number of devices that can be supported by an edge agent, and the maximum number of edge devices that can be supported within their application).
\fi

\iffalse
These can be moved to future work: (1) demonstrating the capability of network systems in handling the traffic produced by the agents of a particular IoT application (to test scaling of a network system in ) and (2) simulating sensor behaviour and (3) mimicking fault behaviour and seeing how analytic engine can identify them (for training to testing IoT fault monitoring tools).

\textbf{MS: Something to be said about the technical design of the language (xtext)}

\textbf{MS: Something to be said about the evaluation.}
\fi

\textbf{Structure.}  Section~\ref{sec:usecase} motivates our  use case for IoT edge-to-cloud simulators. Section~\ref{sec:approach} presents our  DSL. Section~\ref{sec:eval} describes our evaluation of the configurability and usefulness of the DSL. Section~\ref{sec:lesson} highlights industrial adoption and lessons learned. Section~\ref{sec:related} compares with related work. Section~\ref{sec:con} concludes the paper.
%\vspace*{-.5em}
\section{Motivating Use Case}
\label{sec:usecase}
This research resulted from a collaboration with Cheetah Networks (\url{http://cheetahnetworks.com}). Cheetah Networks develops AI-based solutions for real-time monitoring of quality of experience (QoE) in IoT networks. Below, we present the use case that motivated our research. For confidentiality reasons, we do not discuss the use case over our industry partner's products, noting that some of these products are yet to be announced and released. Instead, we use a generic example from the domain of smart cities~\cite{Halegoua:20}; this example has the representative characteristics of IoT edge-to-cloud solutions, which is what our work focuses on.

%IoT edge-to-cloud simulators using an example. We then discuss the need for scalability testing of cloud agents in the context of our example and motivate how such simulators can be used for this purpose. 

In a smart city, sensors monitor a wide range of parameters including, among others, weather, lighting, motion, traffic conditions, utility consumption, water chemistry, and air pollution. The information that these sensors collect is used for decision making and giving commands to actuators that control, for example, traffic and street lights, alarms, barrier gates, pumps, and heating and cooling systems. %Such need can be addressed by developing tailored IoT products. In particular, a product that fits the needs of the above-mentioned sector comprises a number of temperature and humidity sensors that are installed at different locations in a store or storage facility. 
The sensors and actuators that are in close proximity are grouped together and connected in a wired or wireless manner to an edge device, also known as a gateway. %A gateway receives data from the sensors in a periodic, streaming or event-driven fashion. 
Gateways typically have limited computational power, performing only basic data analysis. A gateway sends sensor data -- potentially after some processing -- to cloud services. The communication between a gateway and the cloud is routed through a core network. %The send rate from a gateway to the cloud is typically a configurable parameter. 
Upon receiving unprocessed or semi-processed sensor data, the cloud services perform a more comprehensive processing of the data, generate analytics, determine any necessary course of intervention, and send instructions back to the gateway, which will in turn communicate the instructions to actuators.

Our smart-city example is an instantiation of the IoT architecture in Figure~\ref{fig:edge-cloud}. Within this example, consider the various data storage, data analysis, and decision-making services deployed in the cloud. An important question that the providers of these services need to answer is how far their services will scale in the face of extreme loads from a large number of IoT sensors, actuators, and gateways. Simulation is the most practical strategy for answering this question, noting that a physical IoT testbed for stressing cloud applications is often prohibitively expensive to build.

Creating a reliable simulator nonetheless presents its own challenges. In particular, implementing, within a practical budget, the concurrency and network-based communication required for mimicking a large number of IoT gateways connected to an even larger number of sensors and actuators is a complex technical endeavour. In this paper, we set out to develop an edge-to-cloud simulation DSL that abstracts away from much of the complexity of the underlying concurrency and networking technologies. Through this DSL, we aim to enable software engineers to build reliable edge-to-cloud simulators without requiring extensive knowledge of containerization, virtualization and network protocols.

\section{Edge to Cloud Simulator}
\label{sec:approach}
\lang\ 
is a domain-specific language (DSL) built on top of the conceptual model shown in Figure~\ref{fig:metamodel}. This conceptual model aims to support a scalable and parameterizable architecture for  capturing the communication between edge devices and cloud applications, and to enable simulation-based stress testing of edge-to-cloud solutions.
We explain our conceptual model in Section~\ref{subsec:metamodel}. This is followed by a discussion of \lang\ syntax and usage in Section~\ref{subsec:dsl}.

\subsection{The \lang\ Conceptual Model}
\label{subsec:metamodel}

\begin{figure}
    \centering
    \includegraphics[width=0.9\linewidth]{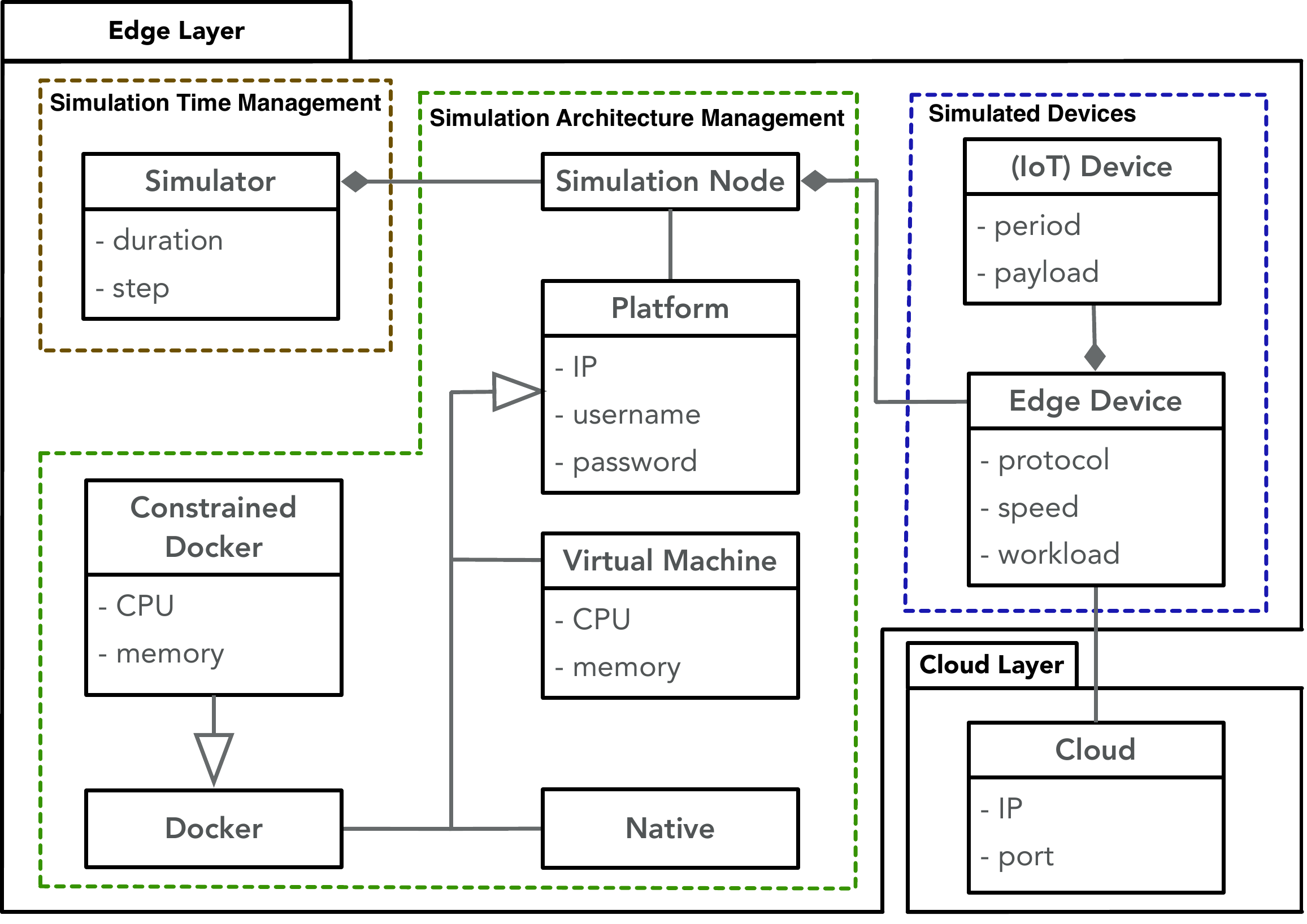}
    \vspace{-.35cm}
    \caption{\lang's underlying conceptual model.}
    \label{fig:metamodel}
    \vspace{-.2cm}
\end{figure}

%Figure~\ref{fig:metamodel} shows the conceptual model underlying the \underline{IoT} \underline{E}dge-to-\underline{C}loud \underline{S}imulation (\lang) framework. 
The concepts in the conceptual model of  Figure~\ref{fig:metamodel} are arranged  under two packages: Edge Layer and Cloud Layer. As described in Section~\ref{sec:intro}, these two layers communicate through a network. The topology of this network is abstracted away in our model since, as argued in Section~\ref{sec:intro}, for stress testing of cloud applications, an ideal network is placed between the edge and cloud layers.  In our model,  network-related information is limited to node IPs, port numbers and  communication protocols.  This information is captured as necessary within the edge and cloud concepts.

%This ideal network can be simulated via a high-throughput switch as we will discuss in Section~\ref{sec:eval}.

%We now elaborate the conceptual model of Figure~\ref{fig:metamodel}.

\textbf{Cloud Layer}. The \texttth{Cloud} concept shown in  Figure~\ref{fig:metamodel} represents a cloud application under stress testing. For an edge-to-cloud simulator to be able to connect to the cloud under test, we need the \texttth{IP} address of the cloud's host and the \texttth{port} at which the cloud receives incoming data. In \lang, we can test several cloud applications by instantiating \texttth{Cloud} multiple times. 

\begin{figure}
    \centering
    \includegraphics[width=.45\textwidth]{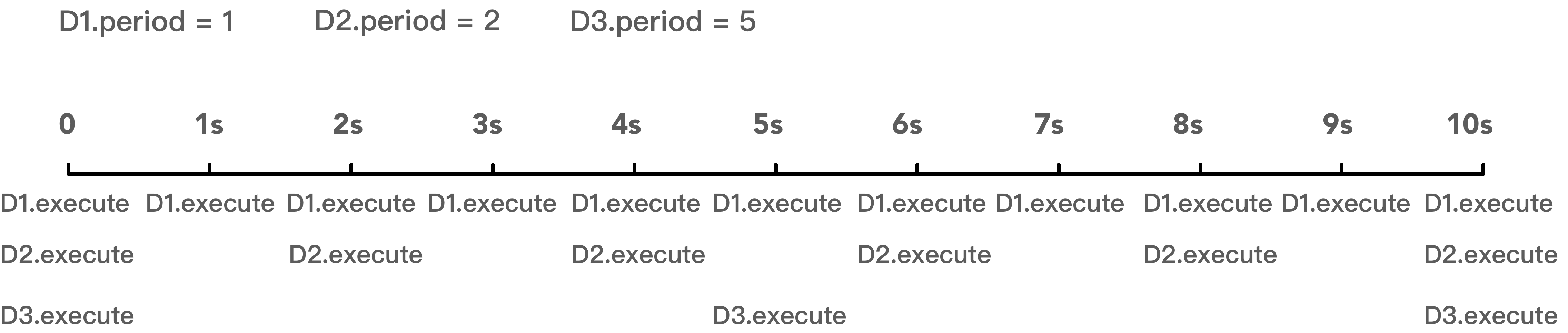}
    \vspace*{-.25cm}
    \caption{An example simulation run representing the execution of three IoT devices, D1, D2, and D3, over a simulation duration of $10$s with each step taking $1$s. Each IoT device regularly executes based on its period.}
    \label{fig:example}
      \vspace*{-.25cm}
\end{figure}

\textbf{Simulation Time Management.} The top-level \texttth{Simulator} concept in Figure~\ref{fig:metamodel} captures the attributes required for managing time. We execute each simulation for a given duration and divide this duration into equal time steps. We use the \texttth{duration} and \texttth{step} attributes to refer to the total simulation duration and to the duration of individual time steps, respectively.
For example, Figure~\ref{fig:example} depicts a simulation run where the  duration is $10$s and each step takes $1$s. 

%\texttth{SimStage} is another important entity in this category. Each \texttth{SimStage} is a time frame of the duration in \texttth{SimulationRun}. With the help of \texttth{SimStage}, the total duration of a single run of simulation is separated into different simulation stages where the settings of simulation are different from each other.

\textbf{Simulated Devices.}  The concepts here are \texttth{IoT Device} and \texttth{Edge Device}. IoT devices encompass sensors (e.g.,  temperature sensors) and actuators (e.g., traffic lights).  IoT devices connect to and communicate periodically with edge devices by wired or wireless links. For example, a temperature sensor may communicate temperature readings to an edge device every $5$s. Edge devices can assume a variety of roles, but their primary function is controlling the data flow at the boundary of a network. Edge devices may perform part of the required processing close to the source (IoT devices) instead of relying on the cloud to do all the processing, which is expensive and can further lead to latency. Edge devices nonetheless usually have limited processing and storage capabilities, and their view on data is limited as well. Edge devices are thus not a substitute for the processing and storage done by the cloud.

Ultimately, what we aim to achieve through \lang\ is high-fidelity simulation of edge devices. For this, we do not need to explicitly simulate IoT devices through individual processes or threads. Neither do we need to distinguish between IoT sensors and actuators, noting that both can have two-way communication with edge devices. Instead, as we elaborate below, we incorporate the relevant properties of IoT devices into the execution of (simulated) edge devices. For succinctness, where there is no ambiguity, we refer to a sensor or actuator simply as ``device''. In contrast, when we mean an edge device, we use the term ``edge device''.

%IoT devices are represented using the \texttth{IoT Device} concept. 
Each instance of \texttth{IoT Device} has a \texttth{period} attribute, indicating the time interval (in terms of the number of simulation steps) between two successive executions of the IoT device. For example, an IoT device with \texttth{period} $= 1$ executes its function every step and the one with \texttth{period} $= 2$ executes every other step (e.g., see Figure~\ref{fig:example}).
Given a set of devices, we choose the size of the step for our simulator as the greatest common divisor of the devices' periods. For example, if we have two  devices, one executing every $4$s and the other every $6$s, then the size of the step for our simulator is set to $2$s. IoT devices, whenever they execute, generate a data packet whose size is specified by the \texttth{payload} attribute.

%Each time an IoT device executes its periodic function,
%to collect the most up-to-date data from its environment, 

%the details of data transmission between IoT devices and edge devices are orthogonal to our work.

\iffalse
Edge devices include edge servers, edge gateways and smartphones that can connect to a network or Internet and have some computing power (often more than devices). Their main purpose is to perform part of data processing close to the source (devices) instead of sending all the data to a remote cloud for processing which is expensive and can lead to latency. However, since the computing power of edge devices is limited, clouds are still needed in \lang.
\fi

%In our model of Figure~\ref{fig:metamodel}, edge devices are represented using the \texttt{Edge Device} concept. 
Each instance of \texttt{Edge Device} owns a group of IoT devices. At every time step, each edge device collects the data packets generated by its associated IoT devices. As mentioned earlier, in  \lang, the IoT device concept does not induce executable entities. Instead, each instance of an edge device, which is an executable entity in \lang, is responsible for emulating the behaviour of the IoT devices associated with it. We provide a precise characterization of the behaviour of edge devices later, when we discuss Algorithm~\ref{algo:edgedevice}.

Edge devices send the collected packets to the cloud following the communication protocol indicated by the \texttth{protocol} attribute in \texttt{Edge Device}. For example, in Figure~\ref{fig:example}, suppose that one edge device is associated with IoT devices D1, D2, and D3. At time 0, the edge device sends the packets from all three to the cloud. At time 1s, it sends the D1's packet, and at time 2s, it sends the packets from D1 and D2 to the cloud. 
In parallel with sending packets, the edge device performs two more operations: (a)~receiving data from the cloud, and (b)~performing compute-intensive operations (i.e., edge computing). Edge devices use a \texttth{workload} attribute to specify the time duration, measured in milliseconds or nanoseconds, spent on compute-intensive operations.

In addition to the \texttth{protocol} and \texttth{workload} attributes, \texttth{Edge Device} has a \texttth{speed} attribute that specifies how many packets an edge device sends to the cloud per simulation time step. To illustrate, let the simulation time step be $1$s. If we set \texttth{speed} to $100$, the edge device sends \emph{up to} $100$ packets over $1$s. That is, the edge device waits for $10$ms after sending each packet to the cloud. In the above, we say ``up to'' because the edge device may send fewer packets if its associated IoT devices produce less packets than what the speed attribute allows to be transmitted. Similarly, if we set \texttth{speed} to $1000$, the edge device waits for $1$ms after sending each packet. To indicate that an edge device can send packets at maximum speed over the time step duration (i.e., without any waiting after sending a packet),  one can set \texttth{speed} to a designated value, \texttt{MAX}.

We use \texttth{speed} to control  the time gap between the sending of the packets. If a large number of (simulated) edge devices attempt to send all their packets to the cloud without any time gap, the simulator may not be able to cope with all the send requests, or these requests may congest the network. In both situations, one may experience packet loss or latency at the (simulated) edge or within the network. In view of our goal, which is to stress test cloud applications, we need to make sure that we do not incur packet loss or latency at the edge or in the network.  Using the speed attribute, we can pace the send requests in a way that would prevent the simulator and the network from being overwhelmed. This in turn allows us to rule out the simulator and the network as the root cause of packet loss and thus localize packet-loss occurrences to the cloud applications. In our empirical evaluation of Section~\ref{sec:eval}, we demonstrate the impact of \texttt{speed} on our results.

%The goal of the simulator is to stress test the cloud and identify the maximum number of edge devices and IoT device that it can handle before the simulator reports that packets are lost. 

\begin{algorithm}[t]
   \footnotesize
\DontPrintSemicolon
\SetSideCommentRight
\SetNoFillComment
    \SetKwBlock{DoParallel}{run in parallel:}{end}
    \SetKwBlock{sendp}{Send Process:}{end}
    \SetKwBlock{receivep}{Receive Process:}{end}
    \SetKwBlock{computep}{Compute Process:}{end}
    \KwIn{\;
    $T$: Simulation  duration\;\\
    $q$: Size of the simulation time step\;\\
    $n$: Number of IoT devices associated to $e$\; \\
    }
    \DoParallel{
       \sendp{
       \For{$i = 0$ to $\frac{T}{q}$}{
         startTime = getWallClockTime() \;
         \For{every $j \in \{1, \ldots, n\}$}{
        
        \If {($i$ \mbox{mod} e.devices[j].period == 0)} 
        {Send e.devices[j].payload to e.cloud using e.protocol\;
        \If {$\mbox{e.speed}$ != MAX } {
        sleep($\frac{\mbox{q}}{\mbox{e.speed}}$) }}
        currentTime = getWallClockTime() \; 
        \If {(q $\leq$ currentTime - startTime)} {
        Break out of the loop (of line~5)\;}}
        %sleep(q - (currentTime - startTime)) \;
        }} 
        \receivep {
        Receive packets from $e.cloud$ \;
        } 
        \computep{
        Perform CPU-intensive operations for duration $e.workload$\; 
        }
    }
    \caption{The behaviour of Edge Device $e$.}
    \label{algo:edgedevice}
\end{algorithm}

Algorithm~\ref{algo:edgedevice} formally specifies the behaviour of edge devices in \lang. An edge device performs three operations in parallel: sending packets, receiving packets and performing (edge) computing. At every time step $i$ between $0$ to $\frac{T}{q}$ where $T$ is the simulation duration and $q$ is the size of the step, the edge device $e$ checks each of its associated IoT devices (line~5). If the period of an IoT device indicates that it should be executed at step $i$ (line~6), then $e$ sends the associated payload to the cloud (line~7). After sending the data, $e$ may wait for some time gap whose size depends on the speed parameter of $e$ (lines~8-9). The loop for sending the IoT devices' data to the cloud (lines~5-12) ends as soon as the time step $q$ has elapsed (line~11-12). As a result, if the time step is too short or too many IoT devices are associated to $e$, some may fail to send their data packets to the cloud.
%in the for-loop on lines 5 to 12. 

The receive method is used for receiving packets transmitted from the cloud to edge device $e$ (lines 13-14). To simulate edge computing in \lang, edge devices perform, for the duration specified by the \texttth{workload} attribute,  some compute-intensive operation (lines 15-16). Note that to properly simulate edge computing, we need to explicitly perform operations that keep the edge CPU busy so that the CPU is not allocated to other parallel tasks. To this end, we perform a series of floating-point operations. 
Algorithm~\ref{algo:edgedevice} stops when the simulation duration $T$ elapses.

%In our conceptual model of Figure~\ref{fig:metamodel}, the concepts of \texttth{IoT Device} and \texttth{Edge Device} map to physical elements present in IoT solutions (e.g., in the use case described in Section~\ref{sec:usecase}). The other elements in  Figure~\ref{fig:metamodel}, however, are only designed for the purpose of building a simulator and do not map to real-world physical elements. 

%At the beginning of each gap in \texttth{SimStage}, \texttth{EdgeAgent} sends one packet for each of the IoT devices belonged to it whose statuses are \texttth{ON} to the \texttth{Cloud} sequentially. For the devices whose statues are \texttth{OFF} in that gap, \texttth{EdgeAgent} does nothing to it. Sizes and protocols of the packets are provided by \texttth{Devices} using \texttth{payload} and \texttth{protocol}. After one packet is sent, \texttth{EdgeAgent} waits \texttth{waitTime} before the next sending for the next \texttth{Device} with \texttth{ON} status as described in \algorithmcfname{1}.

\textbf{Simulation Architecture Management.} An important factor in edge-to-cloud simulation is the ability to simulate a large number of IoT and edge devices. As discussed above, each edge device manages the sending of the packets related to its IoT devices sequentially and within the simulation time step $q$. This assumption matches practice, since in the real-world, edge devices directly handle their related IoT devices. The edge devices themselves, however, run in parallel and are independent from one another. The concepts under Simulation Architecture Management enable a more orderly handling of concurrency, in turn allowing us to optimize the number of edge devices that can run in parallel.

To maximize the capacity of our simulator in terms of the number of parallel edge devices, we introduce an additional tier into our simulator's architecture to group the edge devices into clusters. This tier is represented by the \texttth{Simulation Node} concept in the model of Figure~\ref{fig:metamodel}. Each simulation node is  associated with a platform, represented by the \texttt{Platform} concept. Edge devices that belong to a given simulation node all execute in parallel and on the platform  of the simulation node. The \texttt{Platform} concept has as attributes \texttth{IP}, \texttth{username} and \texttth{password} to enable access to the platform host machine. In \lang, \texttt{Platform} is  specialized into \texttt{Native} (i.e., running on a native operating system), \texttth{Virtual Machine} and \texttth{Docker}. For a virtual machine (VM), we need to specify the \texttth{CPU} and \texttth{memory} attributes, respectively indicating the CPU and memory allocated  to the VM. %In addition, the \texttth{VMIP} attribute is used for setting up each VM and to transfer simulator code to each virtual machine.  
Docker containers can be unconstrained or constrained. For the latter (i.e., \texttth{Constrained Docker}), similar to a VM, we need to specify the \texttth{CPU} and \texttth{memory} attributes.
%for the former, no such parameters are required.  

The idea of clustering edge devices into  simulation nodes in order to scale simulators is inspired by the notion of ``super indivduals'' in massively multi-agent simulators~\cite{Scheffer:95}. 
In our context, such clustering provides a mechanism to reduce contention over CPU, memory and ports. For example, instead of running 100 parallel edge devices on a single computer, we create ten dockers by dividing the single computer resources (CPU, memory and ports) between the dockers equally. We then use each docker to host ten (simulated) edge devices. As we will demonstrate in our evaluation of Section~\ref{sec:eval}, the docker option considerably improves the scalability of our simulator, compared to the native option. 

Another motivation for defining simulation nodes is to be able to restrict the amount of resources used for simulation, thus keeping the cost of simulation within reasonable limits when computation resources need to be purchased from third-parties. We will come back to this in our lessons learned (Section~\ref{sec:lesson}), when we discuss cost-awareness for simulation.

\begin{algorithm}[t]
\footnotesize
\DontPrintSemicolon
\SetSideCommentRight
\SetNoFillComment
\SetKwBlock{DoParallel}{run in parallel:}{end}
    \KwIn{\;
    $m$ : Number of edge devices associated to $\mathit{sn}$}
    \For {every $i \in \{1, \ldots, m\}$} 
    {initialize $\mathit{sn}.edgeDevice[i]$}
    In parallel, run all $\mathit{sn}.edgeDevice[i]$ 
    ($i \in \{1, \ldots, m\}$)  
    \; \tcp*{run Algorithm~\ref{algo:edgedevice} for each $\mathit{sn}.edgeAgent[i]$} 
    \caption{The behaviour of Simulation Node $\mathit{sn}$.}
    \label{algo:simnode}
\end{algorithm}

The behaviour of \texttth{Simulation Node} is specified in Algorithm~\ref{algo:simnode}. Each simulation node, upon creation, initializes each of its associated edge devices (lines 1-2). It then executes the edge devices in parallel (line 3). That is, Algorithm~\ref{algo:edgedevice} is called for each edge device. 

Algorithm~\ref{algo:simulator} shows the overall behaviour of our simulator. The simulator starts by initializing the platforms of the simulation nodes (line~3). It then transfers the simulation code to the platforms (line~4). Next, the simulation nodes start to run in parallel on their respective platforms (line~5). The nodes run until the simulation duration elapses (line~6), at which point the nodes and their platforms are cleaned up (lines~7-9).

Our simulator can be configured to represent different numbers of IoT and edge devices. As these numbers increase, the application(s) in the cloud layer are put under more stress. In this way, our simulator makes it possible to determine how far the cloud application(s) under test can be stretched without degradation in their quality of service. %As we explain further in Section~\ref{sec:eval}, our simulator is executed for a time period. 
Upon the termination of a simulation round, we collect certain metrics, defined in Section~\ref{subsec:metric}. These metrics help determine (1)~whether, given the computational resources made available, the simulator can mimic the desired numbers of IoT and edge devices without getting overwhelmed, and (2)~whether the cloud application(s) are able to process and respond in reasonable time to all the messages \hbox{received from the simulator.}

%(1)~The number of messages  The number of packets that the simulator should have sent versus the number of packets that it actually sent. This measure determines whether the simulator is falling under its own weight or not. (2)~The number of packets that the cloud has received versus the number of packets that it can process.  A number of other statistics is also gathered from the side of the cloud that our simulator is communicating with. In particular,is com executes together with a cloud application , and other statistics that can be gathered using our simulator is the number of messages that are 

%Finally, at the end of the simulation, statistics about the number of sent and received packets and the round-trip transmission time are collected from each \texttth{Platform} and are used for computing average packet loss (\emph{pl}) and average delay (\emph{delay}). These values are returned as output by the simulator (line~6--7). 

\begin{algorithm}[!t]
\footnotesize
\DontPrintSemicolon
\SetSideCommentRight
\SetNoFillComment 
\SetKwBlock{DoParallel}{run in parallel:}{end}
 \KwIn{\;
    \emph{l}: Number of simulation nodes\;\\
    }
% \KwOut{\; 
%   \emph{pl}: average packet loss \;\\
%    \emph{delay}: average delay \;\\
%   }
    
    \For {every $i \in \{1, \ldots, l\}$} 
   {  Let $p$ denote $\mathit{sim}.\mathit{sn}[i].\mathit{platform}$ \;
      Initialize $p$ \;
      Transfer $\mathit{sim}.\mathit{sn}[i]$ to $p$ using $\mathit{p.IP}$, $\mathit{p.username}$ and $\mathit{p.password}$ \;}

    In parallel, run all $\mathit{sim}.\mathit{sn}[i]$ ($i \in \{1, \ldots, l\}$) 
    \;\tcp*{run Algorithm~\ref{algo:simnode} for each $\mathit{sim}.\mathit{sn}[i]$}
    sleep ($\mathit{sim.duration}$) \;
    
    \For {every $i \in \{1, \ldots, l\}$} 
    {
    Terminate $\mathit{sim}.\mathit{sn}[i]$ if it is still running  \;
    Clean up $\mathit{sim}.\mathit{sn}[i].\mathit{platform}$ and $\mathit{sim}.\mathit{sn}[i]$}

   % Compute total packet loss (\emph{pl}) and average time for the transmission of packets from the edge devices to the cloud and back (\emph{delay}) \; 
%    \Return{$pl, delay$}\;
    \caption{The behaviour of Simulator $\mathit{sim}$.}
    \label{algo:simulator}
\end{algorithm}

%In \lang, we can either specify the methods of a cloud entity in detail so that the code of the cloud is auto-generated, or we can assume that the cloud software with which our simulator is expected to communicate already exists and is operational. In the former case, the code of the cloud, after being generated, should be transported and deployed on a machine with \texttt{IP}, \texttt{port}, \texttth{username} and \texttth{password} values indicated as the attributes of the cloud entity in our \lang specification. For the latter case, we need to only indicate the \texttt{IP} and \texttt{port} values of the platform for the existing cloud in our \lang specification. 

\subsection{The \lang\ DSL}
\label{subsec:dsl}
\lang\ aims to support  practitioners in creating simulators that are instances of  the conceptual model of Figure~\ref{fig:metamodel}.  %without having to deal with the implementation details. 
We provide an implementation of \lang\ using Xtext~\cite{Xtext}.
%
%{\bf MS: What I am missing is how an xtext DSL is constructed in general. (1) Describe how a DSL is defined in xtext. Define a conceptual model, define a grammar, do this and that ... State what xtext produces. State the role of xtend (code generation). State how xtend inserts the algorithms of Section 3.1 into the generated simulator code.}
%
%Building on the conceptual model of Figure~\ref{fig:metamodel}, 
Specifically, we use Xtext's grammar language to define \lang's grammar.
%When running the grammar, a new workbench is generated for constructing \lang\ projects. 
To generate executable simulators from specifications written in \lang, we use Xtend~\cite{Xtend}. More precisely, we use Xtend to retrieve all the objects defined in an \lang\ specification; these objects in turn drive the instantiation of a number of a-priori-defined Xtend templates. Our Xtend templates include: (1) Java-based implementations of the algorithms in Section~\ref{subsec:metamodel}, (2) scripts for setting up / starting virtual machines and containers, (3) scripts for uploading and running simulator code on remote platforms, (4) scripts for downloading simulation results from simulation nodes, and (5)~scripts for analyzing and reporting simulation results.

We make \lang's Xtext grammar and Xtend templates publicly available~\cite{IoTECS}. In the rest of this section, and noting that the design of \lang\ follows well-established DSL engineering practices using Xtext, we adopt a practitioner-oriented perspective and emphasize examples of \lang\ usage; full details about the design of \lang\ can be found in our  artifacts~\cite{IoTECS}

%for creating an Xtend code generator which translates the instances of the conceptual model  defined by practitioners in \lang\ project into Java code and bash scripts automatically.
%There are two steps for creating a particular DSL of \lang\  which are model construction step and simulator code translation step. In the first step, practitioners should follow the grammar provided as supplementary material in the Xtext file~\cite{??} to construct their particular instances of  the conceptual model. Then comes to the simulator code generation step. Our implementation translates these models into Java codes using Xtend~\cite{Xtend}. The template for the algorithms of Section~\ref{subsec:metamodel} were implemented in Java and that these templates are instantiated by Xtend during the second step. 
%Below, we present examples of how to use the DSL of \lang\ to construct a simulating system. 

%The syntax of \lang\ is shown in Figure~\ref{??}.We use Backus-Naur Form (BNF) notation~\cite{??} to describe the syntax of \lang: non-terminal symbols are surrounded by angle brackets, while terminal symbols are enclosed in single quotes; ::= means that the symbol on the left must be replaced with the expression on the right; a vertical bar is used to separate alternatives; a star stands for zero or more occurrences; a question mark represents optional elements; parentheses are used for grouping.

\textbf{Cloud.}
Figure~\ref{fig:dslCloudExample} shows a snippet of an \lang\ specification specifying two \texttth{Cloud} instances. For each instance, the \texttth{IP} and \texttth{port} attributes respectively specify the IP address of the cloud application's host machine and the port at which incoming packets are received. We note that the code for instances of \texttth{Cloud} is not meant to be generated by \lang. Rather, each \texttth{Cloud} instance represents an \emph{existing} cloud application that needs to be stress-tested.

\begin{figure}[t]
\begin{lstlisting}[basicstyle=\scriptsize,style=Spec]
Cloud:C1 {
	IP:192.168.0.2
	port:1883
} 
Cloud:C2 {
	IP:192.168.0.3
	port:2605
}
\end{lstlisting}
%\vspace{-.5em}
\caption{Example cloud applications defined using \lang.}
\label{fig:dslCloudExample}
%\vspace{-.3cm}
\end{figure}

% \begin{grammar}
% $\langle\mathit{Cloud}\rangle$ ::=  `Cloud:' $\langle\mathit{ID}\rangle$ `{'  \\
% `IP:' $\langle$IP$\rangle$ \\
% `port:' $\langle$integer$\rangle$ \hfill (3)\\
% (`Methods: {'<cloudMethod> (`,' <cloudMethod>*)?  `}')? `}' \hfill (4)\par
%  <ID> ::= (`a'-`z' | `A'-`Z' | `0'-`9')+ \hfill (5)\par
%  <IP> ::= (`0'-`9')(`0'-`9')?(`0'-`9')?`.'(`0'-`9')(`0'-`9')?(`0'-`9')?`.'(`0'-`9')(`0'-`9')?(`0'-`9')?`.'(`0'-`9')[(`0'-`9')?(`0'-`9')? \hfill  \\
% <CloudMethod> ::= `CloudSend()' | `CloudReceive()' | `CloudCreate('<String>`,'<String>`)' | `CloudStore()' | `CloudCompute()'\hfill(6)
% \end{grammar}

\textbf{Simulator.}
Each \lang\ specification has exactly one instance of the \texttt{Simulator} concept.
%from the model of Figure~\ref{fig:metamodel}.
%
% The syntax for \texttt{Simulator} is:
% \begin{grammar}
% <Simulator> ::= `Simulator:' `{' \hfill (1)\\
% `duration:' <integer> <TimeUnit> \hfill (2)\\
% `quanta:' <integer> <TimeUnit> \hfill (3)\\
% `simulationNodes:{'(<ID>`['<integer>`]') (`,'(<ID>`['<integer>`]')*)?`}'\hfill (4)\\
% `Methods: {' <SimulatorMethod>* `}' `}' \hfill (5)\par
% <TimeUnit> ::= `ms' | `s' | `m' | `h' \hfill (6) \par
% <SimulatorMethod> ::= `analyze()' | `store()'  \hfill (7) 
% \end{grammar}
%
To illustrate, consider the snippet in Figure~\ref{fig:dslSimulatorExample}. 
%The \texttt{duration} variable specifies the duration of simulation and \texttt{step} specifies the time step (as explained in Section~\ref{subsec:metamodel}). 
Here, the simulator has its \texttt{duration} set to 10s and its \texttt{step} to 1s; this results in dividing the simulation into 10 steps of equal length with each step running for 1s. The time unit for \texttt{duration} and \texttt{step} can be in milliseconds (ms), seconds (s), minutes (m) or hours (h). 
A simulator needs to declare its simulation nodes. This is done as illustrated on line~4 of Figure~\ref{fig:dslSimulatorExample}. On this line, we are stating that the simulator has six simulation nodes: five nodes of type \texttt{SN1} and one node of type \texttt{SN2}. We next illustrate how to define the simulation node types (in this case \texttt{SN1} and \texttt{SN2}) and their associated execution platforms.

\begin{figure}[t]
\begin{lstlisting}[style=Spec]
Simulator: {
	duration:10s
	step:1s
	simulationNodes:{SN1[5],SN2[1]}
} 
\end{lstlisting}
%\vspace{-.5em}
\caption{Example simulator defined using \lang.}
\label{fig:dslSimulatorExample}
%\vspace{-.3cm}
\end{figure}

\textbf{Simulation nodes and platforms.} As discussed in Section~\ref{subsec:metamodel}, \lang\ uses the notion of simulation node for grouping simulated edge devices and to further specify the platform on which a group of edge devices run. In the snippet of Figure~\ref{fig:dslHierarchyExample}, we define two simulation node types, \texttt{SN1} and \texttt{SN2}, each having a set of edge devices and a platform. Similar to our convention for specifying the simulation nodes contained in a simulator (see Figure~\ref{fig:dslSimulatorExample}), we define the edge devices contained in a simulation node via an edge-device type and a number of instances. For example, in Figure~\ref{fig:dslHierarchyExample}, \texttth{SN1} has ten edge devices: seven of type \texttth{E1} and three of type \texttth{E2} (we will momentarily illustrate the specification of edge-device types).

Each platform has a type that assumes one of the following  values: \texttth{Native}, \texttth{VM} or \texttth{Docker}. To be able to run a simulation node when its platform is remote (i.e., not the local host), the simulation code and scripts need to be transferred to and set up on the platform first. For this purpose, we need to specify an IP and login credentials (username and password) in instances of the \texttth{Platform} concept. This is illustrated by platform \texttth{P2} (of type \texttth{Docker}) as specified on lines 12-19 of Figure~\ref{fig:dslHierarchyExample}. When the platform is local, this information is not needed, as illustrated by platform \texttt{P1} on lines 9-11 of the same figure. To be precise with respect to our conceptual model of Figure~\ref{fig:metamodel}, for a local platform, the username and password can be viewed as being empty and the IP as 127.0.0.1.
For a \texttth{Docker} platform that is constrained and for any \texttth{VM} platform, we specify the required attributes indicated in our conceptual model of Figure~\ref{fig:metamodel}. For example, platform \texttth{P2} is a constrained docker with \hbox{4 CPUs and 2G of memory.}

% \begin{grammar}
% <SimulationNode> ::= `SimulationNode:' <id> \hfill (1)\\
% `platform:' <id>  \hfill (2)\\
% `EdgeDevices:{'(<ID>`['<integer>`]' [`,'])*`}'\hfill (3)
% \end{grammar}

% The syntax of \texttt{Platform} is:

% \begin{grammar}
% <Platform> ::= `Platform:' <id> `{' \hfill (1)\\
% `type:' <Type>  \hfill (2)\\
% (`locationIP: ' <IP> \hfill (3)\\
% `userName:' <String> \hfill (4)\\
% `password:' <String>)? \hfill (5)\\
% (`cpu:' <integer> \hfill (6)\\
% `memory:' <integer> `G' )? \hfill (7)\\
% (`virtualMachineIP:' <IP>)?  `}'\hfill (8) \par
% <Type> ::= `Native' | `Docker' | `VirtualMachine' \hfill (9) 
% \end{grammar}

\begin{figure}[t]
\begin{lstlisting}[style=Spec]
SimulationNode: SN1 {
	platform:P1
	EdgeDevices:{E1[7],E2[3]}
} 
SimulationNode: SN2 {
	platform:P2
	EdgeDevices:{E1[30]}
} 
Platform: P1{
	type: Native
}
Platform: P2{
	type: Docker
	IP: 192.168.0.4
	username: user2
	password: password2
	CPU: 4
	memory: 2G
}
\end{lstlisting}
\vspace{-.5em}
\caption{Example simulation nodes and platforms defined using \lang.}
\label{fig:dslHierarchyExample}
%\vspace{-.3cm}
\end{figure}

\textbf{Edge and IoT devices.}
Each edge device has a set of IoT devices associated to it. For example, in the snippet of Figure~\ref{fig:dslEdgeDeviceExample}, edge-device type \texttth{E2} is associated with 10 IoT devices of type \texttth{D1} and 20 IoT devices of type \texttth{D2} (we will shortly exemplify IoT-device types).

Each edge device communicates with one cloud application. For example, in Figure~\ref{fig:dslEdgeDeviceExample}, \texttth{E1} is specified as communicating with cloud application \texttth{C1}. The protocol used by an edge device for communication with the cloud is captured by the \texttth{protocol} attribute. Currently, \lang\ supports the \texttth{UDP}, \texttth{TCP} and \texttth{MQTT} protocols. The \texttth{speed} attribute describes the number of packets sent to the associated cloud application in one time step. For instance, in Figure~\ref{fig:dslEdgeDeviceExample}, the \texttth{speed} is set to 100 for \texttth{E1} and 1000 for \texttth{E2}. Since \texttt{step} has been set to 1s (see Figure~\ref{fig:dslSimulatorExample}), \texttth{E1} waits 10ms between sending two consecutive packets; this wait time is 1ms for \texttth{E2}. The \texttth{workload} attribute indicates the amount of edge computing to be done by the edge device (lines 15-16 of Algorithm~\ref{algo:edgedevice}). 
The time unit for \texttt{workload} can be milliseconds (ms), seconds (s) or minutes (m). Note that, in parallel with sending packets and  edge computing, an edge device also receives packets from the cloud (lines 13-14 of Algorithm~\ref{algo:edgedevice}). The protocol for receiving data from the cloud is the same as that used for sending data to the cloud (i.e., is as specified by the \texttth{protocol} attribute).

% \begin{grammar}
% <EdgeDevice> ::= `EdgeDevice:' <ID> `{' \hfill (1)\\
% `protocol:' <Protocol>  \hfill (2)\\
% `speed:' <integer> \hfill (3)\\
% `cloud:' <ID> \hfill (4)\\
% (`devices:{'<ID>`['<integer>`]' (`,'(<ID>`['<integer>`]')*)?`}')?\hfill (5)\\
% `Methods: {' <EdgeMethod>* `}' `}' \hfill (6)\par
% <EdgeMethod> ::= `receive()' | `send()' | `compute()' \hfill(7) \par
% <Protocol> ::= `UDP' | `TCP' \hfill (8) \par
% \end{grammar}

\begin{figure}[t]
\begin{lstlisting}[style=Spec]
EdgeDevice: E1 {
	protocol:TCP
	speed:100
	cloud:C1
	devices:{D1[100]}
} 
EdgeDevice: E2 {
	protocol:TCP
	speed:1000
	cloud:C2
	devices:{D1[10],D2[20]}
	workload:20ms
} 
\end{lstlisting}
\vspace{-.5em}
\caption{Example edge devices defined using \lang.}
\label{fig:dslEdgeDeviceExample}
%\vspace{-.3cm}
\end{figure}

%\texttth{(IoT) Device} is the lowest-level concept modelled within our \texttth{Edge Layer}. 
Each IoT device (type) has two attributes.
The first one is \texttt{period} which determines the frequency of execution. For example, in the snippet shown in Figure~\ref{fig:dslDeviceExample}, device \texttth{D1} generates a packet every second, while device \texttth{D2} generates a packet every two seconds. 
The second attribute is \texttth{payload}. This attribute can be used to specify the actual payload content, e.g., \texttth{payload: "23C"}, or alternatively, the size of the packet that the device sends every time the device executes. In the latter case, the payload unit can be bytes (B), kilo bytes (kB) or mega bytes (MB). When a size unit is indicated for the payload, the content of the payload is generated randomly as per the requested size. This option, illustrated in Figure~\ref{fig:dslDeviceExample}, is convenient when the actual payload is unimportant for simulation purposes (e.g., in a simple cloud storage application).

% \begin{grammar}
% <Device> ::= `Device:' <ID> `{' \hfill (1)\\
% `period:' <integer>  \hfill (2)\\
% `payload:' <integer> <PayloadUnit> `}' \hfill (3)\par
% <PayloadUnit> ::= `b' | `Kb' | `Mb' \hfill (4) 
% \end{grammar}

\begin{figure}[t]
\begin{lstlisting}[style=Spec]
Device: D1 {
	period:1
	payload:60B
} 
Device: D2 {
	period:2
	payload:100B
} 
\end{lstlisting}
\vspace{-1em}
\caption{Example (IoT) devices defined using \lang.}
\label{fig:dslDeviceExample}
\vspace{-1em}
\end{figure}

\section{Evaluation}
\label{sec:eval}
In this section, we evaluate the applicability and usefulness of simulators generated from \lang\ specifications for stress testing of cloud applications. We use the term \emph{simulator} to refer to the executable code generated from instantiating the edge-layer concepts of the model of Figure~\ref{fig:metamodel}; we use the term \emph{cloud} to refer to instances of the cloud concept in this model. For our experiments, as we  discuss in Section~\ref{subsec:cloudBL}, we create our own baseline cloud applications. 
The research questions (RQs) that we investigate are as follows:

\textbf{RQ1. (Configuring Simulators)} \emph{Can we configure our simulator so that it can successfully simulate a large number of IoT and edge devices?} With RQ1, we examine whether the simulator architecture envisaged by the conceptual model of Figure~\ref{fig:metamodel} allows one to instantiate a large number of IoT and edge devices as required by our motivating use case (Section~\ref{sec:usecase}). For practical reasons, we want the whole simulator to run successfully on a single computer with modest resources. In RQ1, we  assess how grouping edge devices into simulation nodes and executing simulation nodes on alternative  platforms (native, virtual machine and docker) impacts the ability of simulators to scale. The results of RQ1 lead to a configuration range for simulators in terms of the number of IoT devices, edge devices and simulation nodes as well as the platform types. %that can successfully run on a host machine. 
%and communicate a large number of messages to the cloud. 

\textbf{RQ2. (Cloud-Application Stress Testing)} \emph{Can our simulator be used for stress testing cloud applications and determining the number of IoT and edge devices that a cloud application can handle?}  We perform simulations using the  configurations identified in \emph{RQ1} to assess how well one can push a cloud application to its limits in terms of the number of IoT and edge devices being handled.

\iffalse
In particular, we create alternative cloud instances with different intensities of computing versus IO tasks by introducing compute-intensive operation cycles in instances of the cloud concept varying the workload attribute in instances of the cloud concept (see Figure~\ref{fig:metamodel}). We then compare the simulation results for different cloud instances to demonstrate that the changes in the simulation results are consistent with the intensity of IoT versus computing activities. 
\fi

\subsection{Baseline Cloud Applications}
\label{subsec:cloudBL}
%\begin{algorithm}[t]
%   \footnotesize
%\DontPrintSemicolon
%\SetSideCommentRight
%\SetNoFillComment
%    \SetKwBlock{DoParallel}{run in parallel:}{end}
%    \SetKwBlock{sendreceivep}{Send and Receive Process:}{end}
%    \SetKwBlock{computep}{Compute Process:}{end}
    
%    \If {$c.username$ != null and $c.password$ != null}{Access the $c.IP$ machine using $c.username$ and $c.password$ to transport the code of Cloud to $c.IP$\;
%    Start Cloud \;
%    } 
%        \While{true}{
%           Receive a packet from an edge device\;
%           Perform CPU-intensive operations for the duration \emph{cloudComuting}\;
%           Send the packet back to the same edge device\;
%        }
%   \caption{The Behaviour of our baseline Cloud.}
%    \label{algo:cloud}
%\end{algorithm}

%In Section~\ref{sec:lesson}, we will present qualitative lessons from applying \lang\ in collaboration with our partner, Cheetah Networks. Nevertheless, 
Due to confidentiality, for RQ1 and RQ2, we do not report quantitative  results over our partner's cloud applications. Instead, we build our own \emph{baseline cloud applications} and apply to these baselines the same analysis as that performed on our partner's applications. The  results we report for RQ1 and RQ2 are over the baselines.
%The behaviour of the baseline cloud application is shown in Algorithm~\ref{algo:cloud}. 
Our baselines implement a simple loop where an instance of the cloud concept receives a packet from an edge device, performs compute-intensive  operations for a configurable duration, \emph{cloudComputing}, and  sends the received packet back to the sender. The compute-intensive operations are meant at simulating cloud computing. We create three baselines, varying the \emph{cloudComputing} parameter to $0$, $1$ms, and $5$ms, respectively.

%We assume that if \texttth{username} and \texttth{password} attributes are set to null the cloud application is already deployed on its host machine;

%On the cloud side, we capture a single entity type representing a cloud application. In \lang, we can create several instances of the cloud entity at different locations. The \texttth{cloud} entity has an \texttth{IP} address indicating the address of the host computer for the cloud application. The \texttth{port} attribute determines which port of the cloud host machine is used for receiving incoming traffic from the edge layer. The \texttth{username} and \texttth{password} attributes are used to access the cloud host computer to upload the cloud software code. Generally speaking, cloud applications perform two kinds of tasks: (1)~I/O operations by receiving packets from the edge layer and sending information back to the edge layer, and (2)~CPU-intensive or cloud computing operations to process the data received from the edge layer. We use two attributes, \texttth{inToOut} and \texttth{workload}, to configure these two tasks for a cloud entity. The \texttth{inToOut} attribute determines the ratio of the messages received to the messages sent, and \texttth{workload} determines the time duration that the cloud entity spends on performing compute-intensive operations for each I/O task.

\subsection{Experiment Design}
\label{subsec:expdesign}
To answer RQ1 and RQ2, we use two computers, one for running the simulator and one for running a  baseline cloud application.

\begin{figure}
    \centering
    \includegraphics[width=.7\linewidth]{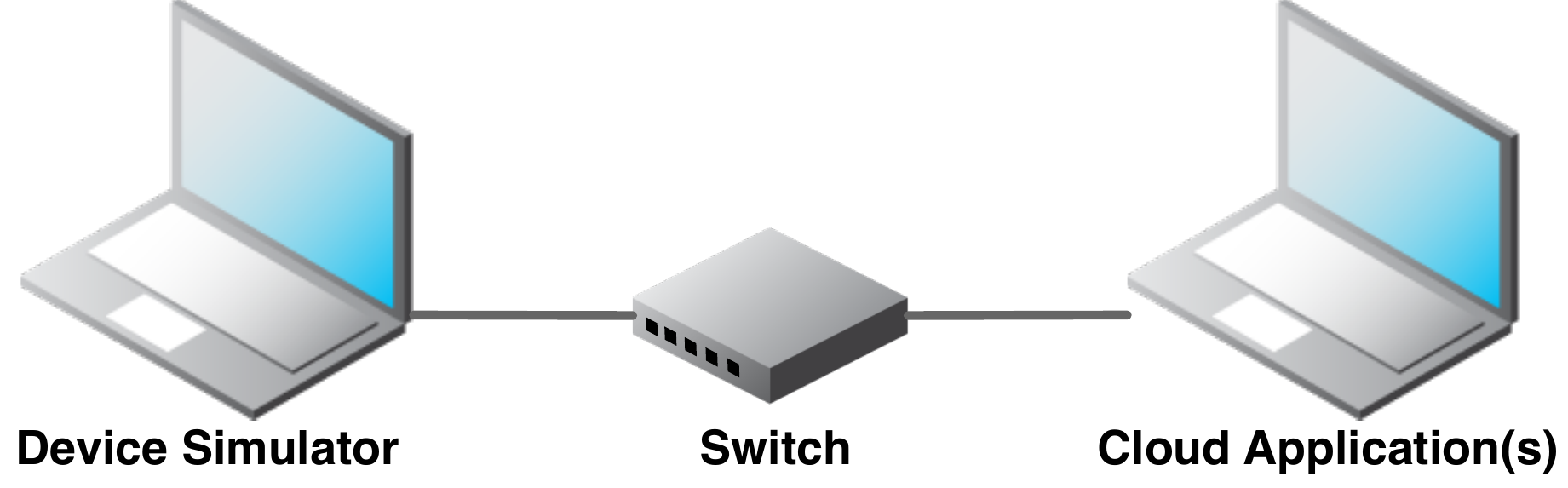}
    \vspace*{-.2cm}
    \caption{Physical setup for our evaluation.}
    \label{fig:evalsetup}
     \vspace*{0em}
\end{figure}

Consistent with our objective of abstracting away from the complexities of networks (discussed in Section~\ref{sec:intro}), we connect the simulator to the cloud via a single switch as shown in Figure~\ref{fig:evalsetup}. This setup, while simple, provides a high-fidelity testbed for stress testing of cloud applications, as it routes all network traffic through the switch (ideal network) rather than a real-world network that may not be as reliable or predictable. Requiring the simulator to run on a single machine (laptop) was in line with the needs of our partner; they needed the simulator machine to be portable so that it could be brought to different sites and connected to different networks.

%In Section~\ref{sec:lesson}, we discuss some important considerations in the context of our collaboration with Cheetah Network to ensure the fidelity of our simulation setup.

 %Since our goal is to test the cloud instead of modelling or characterizing the edge layer, we run the entire simulator on a single machine and the cloud on another machine. 
 \iffalse % future work
 We note that our simulator can also be used for other use cases such as testing the behaviour of IoT devices and edge devices. For those use cases, however, the simulator cannot be executed on a single machine and has to be executed on a distributed setup. Investigating such use cases are left for future work. 
\fi

Table~\ref{tab:parameters} shows the configuration parameters for our experiments. We select the duration of simulations to be $2$s and the time step to be $0.5$s (i.e., four steps in each simulation). For the device payload size, we follow the recommendation of our collaborating partner, and set it to $8B$.
We set the period of each IoT device to 1 to maximize the number of messages sent in each time step. Based on the recommendation of domain experts, we configure virtual machines and constrained dockers with four CPU cores and $2$G of memory.  For edge devices, we eliminate edge-computing time by setting \texttth{edge.workdload} to zero, noting that our goal is maximizing communication with the cloud. For the number of IoT devices per edge device and the number of edge devices per simulation node, we followed the recommendation of our partner based on past examples of large-scale IoT rollouts.

\begin{table}
\caption{Parameters of our experiments.}
\label{tab:parameters}
\vspace*{-.5em}
\begin{center}
\scalebox{.65}{
\begin{tabular}{p{3.2cm} p{0.6cm}||p{3.8cm} p{3cm}}
\toprule
duration &  2s & cloud computing time & 0/1/5ms  \\
\hline
 time step   &  0.5s &   \# of sim nodes &  8 to 12\\
\hline
\# of IoT devices per edge device &  100 & \# of edge devices per sim node  &  10 \\
\hline
(IoT) Device.period &   1  & (IoT) Device.payload &  8B \\
\hline
Edge Device.workload &   0 &Edge Device.speed & Max, 500, 250, 167, 125 \\
\hline
 Virtual Machine.CPU & 4 & Constrained Docker.CPU & 4  
\\
\hline
 Virtual Machine.memory & 2G  & Constrained Docker.memory & 2G
\\

\bottomrule
\end{tabular}}
\vspace*{-.35cm}
\end{center}
\end{table}

To answer \textbf{RQ1}, we executed our simulator using the parameters of Table~\ref{tab:parameters}. Specifically, for the speed of edge devices, we insert a time gap of $0$ms, $1ms$, $2ms$, $3ms$ and $4ms$ between the consecutive messages sent, resulting in the edge device speed to be set, respectively, to MAX, $500$, $250$, $167$ and $125$.  Note that further decreasing the speed (i.e., increasing the gap time) leads to inconsistency in our experimental setup since each edge device needs to be able to send 100 messages in a time step (i.e., one message per IoT device).
For the number of simulation nodes, the partner was interested in having ten. Nonetheless, the partner was also interested in determining how far the simulation could be pushed on the laptop that we were using without the simulator showing anomalous behaviours. Based on this requirement, we experimented with 10 $\pm$ 2 simulation nodes, thus the range of $8$ to $12$ nodes in Table~\ref{tab:parameters}.  With these numbers of simulation nodes, we use our simulator to capture between $8000$ to $12000$ IoT devices where each is sending a message every half second to the cloud. We executed each configuration of our simulator for four platform options: Native, Virtual Machine (VM), Constrained Docker (CDC) and Unconstrained Docker (UDC).   We assumed that the cloud is not doing any cloud computing (i.e., \emph{cloud computing} = 0), since the purpose of \textbf{RQ1} is configuring the simulator. This in turn requires operating under the assumption that the cloud responds as promptly as possible. In total, to answer \textbf{RQ1}, we performed 100 experiments (\# of platforms = 4) $\times$ (\# of speeds = 5) $\times$ (\# of sim nodes = 5). We repeated each experiment 10 times to account for random variation.

To answer \textbf{RQ2}, we consider the simulation configurations that, based on the results of $\textbf{RQ1}$, can successfully execute on a single machine and send all messages to the cloud. We then rerun these configurations for two cloud instances where their computing time is set to $1$ms and $5$ms, respectively. That is, in contrast to the cloud instance used for \textbf{RQ1}, here, we consider cloud instances that have to preform some computing tasks after receiving messages and before responding. For \textbf{RQ2}, we perform 100 experiments  (\# of platforms = 2) $\times$ (\# of speeds = 5) $\times$ (\# of sim nodes = 5) $\times$ (\# of alternative cloud instances = 2) and repeat each experiment $10$ times to account for randomness. Note that, for RQ2, we consider the CDC and UDC platforms only. This is because the VM and Native platforms could not pass the sanity check in RQ1: in our experimental setup, these platforms could not successfully run the simulator without incurring packet loss. We executed our simulator on a machine with a 2.5 GHz Intel Core i9 CPU and 64 GB of memory, and our cloud on a machine with a 2.3 GHz Intel Core i9 CPU and 32 GB of memory. We connected the two machines using an unmanaged NETGEAR GS308v3 Gigabit Ethernet switch.

\textbf{Implementation and availability.} We implemented \lang\  using Xtext (version 2.25.0)~\cite{Xtext} and Xtend~(version 2.25.0)~\cite{Xtend}. As noted in Section~\ref{subsec:dsl}, the \lang\ grammar and our code generation tool are publicly available~\cite{IoTECS}. 

%Xtext is used to define our grammar of \lang\ while Xtend is used as the code generator which translates codes in \lang\ to Java. 

\subsection{Metrics}
\label{subsec:metric}
We record two main metrics as the results of our experiments: (1)~the number of packets that are dropped during the simulation, and (2)~the packet transmission time, i.e., the amount of time it takes for the packets to be transmitted from the simulator to the cloud and be processed by the cloud. In the setup of Figure~\ref{fig:evalsetup}, packets can be dropped (1) on the simulator side when the simulator fails to send all the messages it is expected to send, (2) on the switch (network) side when the switch fails to handle the traffic that needs to pass through it, e.g., due to congestion, or (3) on the cloud side, when the cloud fails to receive all the packets that have reached its host computer, or fails to send responses for the packets it has received. In our experiments, we measure packet drops at the three above locations separately and refer to them respectively as \emph{SimDrop}, \emph{NetDrop} and \emph{CloudDrop}. Note that in order to measure these values, we have to keep track of the packet counts at the level of the simulator, the cloud and also the network adapters on both host machines. To count the number of packets sent from and received by the network adapters of the host machines of the simulator and the cloud, we use the Wireshark tool~\cite{Wireshark} which is the world's most widely-used network protocol analyzer. Using Wireshark, we have confirmed that, in our experiments,  the packet drop at the network adaptors is zero or negligible (less than five packets out of hundreds of thousands). In addition,  since  we connect the computers using a switch (whose capacity we do not exceed) the packet drop for the network is zero as well. Hence, we  do not report \emph{NetDrop} as it is always zero for our experiments.
The packet drop values measured at the simulator (i.e., \emph{SimDrop}) and at the cloud (i.e, \emph{CloudDrop}) as well as \emph{packet transmission time} (as defined above) are thus the only measures that we need to report in order to capture the behaviour of the simulator and the cloud.

Among these measures, \emph{SimDrop} determines if the simulator is able to scale and that it does not fail under its own load. If there is packet drop by the simulator, it means that the simulator is overwhelmed. The \emph{CloudDrop} and the packet transmission time (\emph{TransTime}, for short) determine how well the cloud is able to handle the messages it receives from the edge.

\begin{table*}[t]
  \caption{Average packet drop and  transmission time values obtained by ten runs of the experiments of RQ1.}
  \vspace{-1em}
  \label{tbl:rq1}
{ \fontsize{4.8}{5.5}\selectfont
  \setlength\tabcolsep{1pt} % default value: 6pt
  \begin{tabularx}{\columnwidth*2}{X X X X X X X X X X X X X X X X X X X X X X X X X X X}
    \toprule
\textbf{Metrics} &&\textbf{\#SimulationNode=12} &&&&&\textbf{\#SimulationNode=11} &&&&& \textbf{\#SimulationNode=10}&&&&&\textbf{\#SimulationNode=9} &&&&& \textbf{\#SimulationNode=8} \\
Speed&&MAX&500&250&167&125&MAX&500&250&167&125&MAX&500&250&167&125&MAX&500&250&167&125&MAX&500&250&167&125\\
    \midrule
&&&&&&&&&&&&&\textbf{CDC}&&&&&&&&&&&&&\\
SimDrop&&0&0&0&0.2&0&0&0&0&0&0&0&0&0&0&0&0&0&0&0&0&0&0&0&0&0\\
%NetLoss&&0&0&0&0&0&0&0&0&0&0&0&0&0&0&0&0&0&0&0&0&0&0&0&0&0\\
CloudDrop&&2314.5&0&0&0&0&3246.1&70.5&0&0&0&1517.2&0&0&0&0&768&0&0&0&4.3&0&0&0&0&0\\
TransTime(ms)&&6.61&0.49&0.74&0.52&0.29&10.57&4.31&1.32&0.33&0.28&9.01&1.74&0.70&0.29&0.27&4.71&1.38&0.62&0.28&0.62&0.99&0.98&0.35&0.28&0.26\\
 &&&&&&&&&&&&&\textbf{UDC}&&&&&&&&&&&&&\\
SimDrop&&0&0&0&0&0&0&0&0&0&0&0&0&0&0&0&0&0&0&0&2.2&0&0&0&0&0\\
%NetLoss&&0&0&0&0&0&0&0&0&0&0&0&0&0&0&0&0&0&0&0&0&0&0&0&0&0\\
CloudDrop&&2805.5&0&0&0&0&3741.2&0&0&0&0&296.6&0&0&0&0&0&0&0&0&0&0&0&0&0&0\\
TransTime(ms)&&7.05&0.48&0.61&0.46&0.31&8.11&1.30&0.87&0.37&0.30&3.52&1.00&0.61&0.34&0.29&0.62&0.61&0.51&0.30&0.28&0.41&0.49&0.41&0.29 & 0.27\\
  &&&&&&&&&&&&&\textbf{VM}&&&&&&&&&&&&&\\
SimDrop&&42.6&3254.4&2922.2&4383.9&6294.2&71&119.10&1272.9&2407.3&4410.3&69.1&423.7&178.3&710.1&2322.9&0&107.7&79.6&50.4&844.8&26.8&0.9&0&0&100.7\\
%NetLoss&&0&0&0&0&0&0&0&0&0&0&0&0&0&0&0&0&0&0&0&0&0&0&0&0&0\\
CloudDrop&&28.1&0&0&0&0&0&0&0&0&0&0&0&0&0&0&0&0&0&0&0&0&0&0&0&0\\
TransTime(ms)&&0.47&0.44&0.27&0.28&0.28&0.38&0.28&0.28&0.28&0.28&0.31&0.28&0.28&0.50&0.28&0.29&0.26&0.26&0.30&0.28&0.27&0.27&0.27&0.26&0.30\\  
  &&&&&&&&&&&&&\textbf{Native}&&&&&&&&&&&&&\\
SimDrop&&25051.1&25563.8&25958.1&26089.8&26312.4&21022.4&21543.4&21980.3&22060.8&22599.6&17252.6&17546.1&18103&18128.8&18537&13239.5&13834&13951.9&14209.4&14136.1&9323.7&9858.3&9935.7&10176.4&10168\\
%NetLoss&&0&0&0&0&0&0&0&0&0&0&0&0&0&0&0&0&0&0&0&0&0&0&0&0&0\\
CloudDrop&&0&0&0&0&0&0&0&0&0&0&0&0&0&0&0&0&0&0&0&0&0&0&0&0&0\\
TransTime(ms)&&0.35&0.26&0.26&0.26&0.26&0.26&0.26&0.28&0.33&0.30&0.26&0.26&0.29&0.32&0.26&0.26&0.28&0.28&0.26&0.26&0.26&0.26&0.30&0.26&0.29\\  
    \bottomrule
  \end{tabularx}
}
\end{table*}

%As discussed in Section~\ref{subsec:expdesign}, the purpose of \textbf{RQ1} is to identify simulator configurations that can operate without incurring any packet loss on the side of the simulator. For this purpose, we eliminate cloud computing and run our simulator using our four different platform options, i.e., native, VM, constrained docker (CDC) and unconstrained docker (UDC), while varying the number of simulation nodes and the speed of edge devices based on the values specified in Table~\ref{tab:parameters}.

\subsection{Results}
\textbf{RQ1.}
Table~\ref{tbl:rq1} shows the average values for SimDrop, CloudDrop, and packet transmission time obtained by running our simulator 
using the parameters in Table~\ref{tab:parameters}. Here,  cloud computing is fixed at zero, i.e., the cloud does not perform any computing (beyond sending and receiving packets).  As discussed in Section~\ref{subsec:expdesign}, for RQ1, we run our simulator using four different platform options, i.e., Native, VM,  CDC and UDC. In the native option, we run all the edge devices in parallel on the simulator host machine without grouping them into separate simulation nodes. In contrast, when we use VM, CDC, UDC, we group every ten edge devices into a simulation node and run each simulation node on a separate VM, CDC or UDC. For each VM and each CDC, we use the CPU and memory sizes specified in Table~\ref{tab:parameters}.

As shown in Table~\ref{tbl:rq1}, there is substantial packet drop on the simulator side when we use the Native option. This indicates that in this architecture where edge devices are not grouped into simulation nodes, the simulator is simply unable -- due to resource contention -- to send all the packets it is supposed to send. This highlights the importance of the hierarchical structure provided by the \texttth{Simulation Node} concept in \lang. Similarly, the VM option, while generating SimDrop values lower than those of the Native option, is unsuitable for stress testing as it still leads to non-negligible packet drop on the side of the simulator.

For the simulation configurations executed on CDC and UDC, SimDrop values are zero (or negligible). For these configurations, the cloud side drops packets, particularly  when edge devices send their packets  at once and without having any time gap in between (i.e., when the edge device speed is MAX). By inserting a time gap on the side of edge devices (i.e., lowering the speed), the cloud will receive the packets at a lower rate and can avoid  packet drop.

Note that transmission time values for the Native and VM options should be discarded. This is because, for these options, there is high packet drop at the simulator, and hence, considerably fewer packets are transmitted from the simulator to the cloud and vice versa. Therefore,  the transmission times are computed for a smaller portion of packets, leading to unrealistically small values. For the CDC and UDC options where the simulator is able to complete its task without packet drop, the transmission time decreases as we reduce the speed of the edge devices. This trend is expected, since by reducing the speed, we avoid the cloud side from being overwhelmed, hence reducing the packet transmission time.

%the packet congestion happened on Cloud side is gradually reduced.  Like CloudLoss, the Delay values of DCC and DCU are bigger than VM when Speed is high and the values keep decreasing when slowing down the sending speed until they are lower than 0.3 milliseconds which is the approximate delay time when there is no congestion. As there is no congestion for Native, the values of Delay stay around 0.3 milliseconds regradless of the number of SimulationNodes and sending speed. The results signify the need for speed adaptation for DCC, DCU and VM. 

 \resq{The answer to \textbf{RQ1} is that  our simulator can be tuned such that it can, on a conventional laptop, simulate the sending of $24000$ packets per second without packet drop. Further, our results show the importance of grouping edge devices into simulation nodes to manage resource usage on the simulator's host machine and thus enable the simulator to scale as much as possible. In our experimental setup,  containerization (through dockers) turned out to be the best option for scaling. Among the main benefits of \lang\ are its support for simulation nodes and its ability to abstract from platform details so that engineers can explore different platform options.}

\begin{table*}[t]
  \caption{Average packet drop and  packet transmission time values obtained by ten runs of the experiments of RQ2.}
  \vspace{-1em}
  \label{tbl:rq2}
{ \fontsize{4.8}{5.5}\selectfont
  \setlength\tabcolsep{1pt} % default value: 6pt
  \begin{tabularx}{\columnwidth*2}{X X X X X X X X X X X X X X X X X X X X X X X X X X X}
    \toprule
\textbf{Metrics} &&\textbf{\#SimulationNode=12} &&&&&\textbf{\#SimulationNode=11} &&&&& \textbf{\#SimulationNode=10}&&&&&\textbf{\#SimulationNode=9} &&&&& \textbf{\#SimulationNode=8} \\
Speed&&MAX&500&250&167&125&MAX&500&250&167&125&MAX&500&250&167&125&MAX&500&250&167&125&MAX&500&250&167&125\\
    \midrule
 &&&&&&&&&&&&&\textbf{CDC\_Cloud1ms}&&&&&&&&&&&&&\\
%SimLoss&&0&0&0&0&0&0&0&0&0&0&0&0&0&0&0&0&0&0&0&0&0&0&0&0&0\\
%NetLoss&&0&0&0&0&0&0&0&0&0&0&0&0&0&0&0&0&0&0&0&0&0&0&0&0&0\\
CloudDrop&&45982.9&45924.2&45893.6&45787.1&45664.5&42013.7&41949.3&41886.3&41760.7&41662.9&38029.8&37977.8&37930.2&37809.7&37679.2&34047.8&33994.5&33937.8&33806.5&33700.7&30092.1&30062.6&29966.5&29839.2&29686.2\\
TransTime(ms)&&744.6&749.4&753.6&744.3&735.6&748.7&754.6&743.7&736.5&735.1&747.1&747.9&748.1&744.4&737.8&748.6&748.4&743.8&743.1&743.4&743.3&744.8&742.7&746.4&730\\  
 &&&&&&&&&&&&&\textbf{CDC\_Cloud5ms}&&&&&&&&&&&&&\\
%SimLoss&&0&0&0&0&0&0&0&0&0&0&0&0&0&0&0&0&0&0&0&0&0&0&0&0&0\\
%NetLoss&&0&0&0&0&0&0&0&0&0&0&0&0&0&0&0&0&0&0&0&0&0&0&0&0&0\\
CloudDrop&&47021&47007.3&46997.6&46978&46950.8&43025.1&43011.3&43003.1&42981&42948.9&39028.3&39017.5&39006.6&38978.6&38953.1&35031.7&35018.6&35009.8&34982.4&34955.3&31036&31029.4&31011.6&30984.5&30962.4\\
TransTime(ms)&&2899.2&2877.2&2865&2878.6&2832.7&2892.2&2864.4&2869.5&2854.4&2809.2&2873.1&2873.7&2873&2824.8&2808.8&2855.9&2859.1&2851.8&2831.2&2834&2879.8&2867.1&2854.5&2845.6&2849.2\\  
 &&&&&&&&&&&&&\textbf{UDC\_Cloud1ms}&&&&&&&&&&&&&\\
%SimLoss&&0&0&0&0&0&0&0&0&0&0&0&0&0&0&0&0&0&0&0&0&0&0&0&0&0\\
%NetLoss&&0&0&0&0&0&0&0&0&0&0&0&0&0&0&0&0&0&0&0&0&0&0&0&0&0\\
CloudDrop&&46046.9&45970.7&45943.6&45754.1&45634.9&42035.9&42039.6&41950.5&41795.5&41665.9&38051.3&38061.3&37976.3&37819.1&37682.6&34055.8&34060.5&33951.9&33810.2&33706.6&30076.2&30066.4&29969.5&29823.5&29759.7\\
TransTime(ms)&&736.8&739.1&744.4&711.9&722.1&733.1&758.7&744.4&734.6&728.2&736.5&757.2&749.6&740.4&734.7&741.1&739.6&740.5&738.1&738.1&736.7&738.8&736.6&735.7&752.2\\  
 &&&&&&&&&&&&&\textbf{UDC\_Cloud5ms}&&&&&&&&&&&&&\\
%SimLoss&&0&0&0&0&0&0&0&0&0&0&0&0&0&0&0&0&0&0&0&0&0&0&0&0&0\\
%NetLoss&&0&0&0&0&0&0&0&0&0&0&0&0&0&0&0&0&0&0&0&0&0&0&0&0&0\\
CloudDrop&&46998.7&47005&46997.7&46978.1&46954.1&43024.7&43013.4&42999.7&42979.9&42954.1&39027&39018.5&39006&38981.5&38953&35028.4&35026.8&35010.5&34986.1&34955.3&31030.3&31029.7&31014.1&30983.1&30962.9\\
TransTime(ms)&&2873&2882.3&2848.2&2866.1&2830.4&2876.5&2854.6&2863.7&2847.4&2839.9&2867.4&2866.8&2871.3&2857.7&2831.6&2851.2&2856.6&2867.5&2850.5&2827.3&2864.6&2859.8&2864.3&2821.3&2846.5\\  
    \bottomrule
  \end{tabularx}
}
\end{table*}

\textbf{RQ2.}
To answer \textbf{RQ2}, we create two new instances of our cloud by setting the cloud computing variable to $1$ms and $5$ms, respectively. We refer to the former instance as Cloud1ms, and to the latter as Cloud5ms. We then apply the CDC and UDC configurations of our simulator from RQ1 to stress test Cloud1ms and Cloud5ms. These instances represent more realistic behaviours from the cloud compared to the cloud instance in RQ1 which had zero computing and was intended at tuning the simulator.  

Table~\ref{tbl:rq2} shows the CloudDrop and transmission time values obtained by applying the CDC and UDC configurations of our simulator to Cloud1ms and Cloud5ms. We note that the SimDrop values for this experiment were all zero and hence are not shown in the table. The results of Table~\ref{tbl:rq2} as well as those for the cloud with no computing in Table~\ref{tbl:rq1} (CDC and UDC cases) show that, by increasing the computation on the cloud side, the ability of the cloud in receiving and sending messages  decreases. This is reflected by higher CloudDrop and packet transmission time values. %The results in these two tables further show that the cloud instance with zero computing is able to receive messages from a maximum of 80 edge devices (or 8000 IoT devices) without incurring packet loss. In order to increase its capacity to handle more IoT and edge devices, we need to reduce the speed of edge devices. In addition, by introducing computing in our cloud instances, it can no longer manage to receive messages from the same number of edge devices and devices. 

%In Section~\ref{sec:lesson}, we discuss our experience applying our simulator for cloud stress testing in industry. 

 \resq{The answer to \textbf{RQ2} is that as  cloud-computing time increases, so does the cloud's packet drop and packet transmission time. This means that our simulator, without incurring any packet drop on its end, has successfully fulfilled its function, i.e., pushing the cloud application under test to its limits.}

\textbf{Threats to Validity.} Construct and external validity are the most relevant dimensions of validity for our evaluation.

\emph{Construct Validity:} The main threat to construct validity is whether our evaluation metrics, as defined in Section~\ref{subsec:metric}, can be measured reliably. To this end, we note that our metrics are based on common practices in network and performance engineering. To accurately measure our metrics of interest, we used Wireshark~\cite{Wireshark} -- a standard and widely used network protocol analyzer.

\emph{External Validity:} We have used \lang\ to generate simulators for testing our baseline cloud applications (Section~\ref{subsec:cloudBL}) as well as industrial cloud applications at Cheetah Networks. Due to confidentiality, we did not report results for our partner's applications. The process we followed for stress testing these industrial applications is the same as that presented in this paper. Furthermore, \lang\ is currently being used independently by our partner (see Section~\ref{sec:lesson}). The above said, applying \lang\ in other industrial contexts remains necessary for further improving external validity.
\section{Adoption and Lessons Learned}
\label{sec:lesson}
In this section, we report the adoption status of \lang\ and outline the lessons we learned from the development of this DSL.

\subsection{Uptake by Industry Partner} 
\lang\ was developed in collaboration with Cheetah Networks over a period of seven months, spanning September 2021 to February 2022. During this period, we iteratively improved \lang\ in response to feedback from the experts at Cheetah. \lang\ has since been used independently by Cheetah in several of  their testing and demonstration campaigns. %%As already noted in Section~\ref{sec:usecase}, due to confidentiality, we cannot disclose the specifics of the industrial test subjects at Cheetah. %We can nonetheless disclose that these test subjects are cloud components that feed information into Cheetah's IoT network anomaly detection engine, ARTINA (Actionable Real-Time IoT Network Analytics)~\cite{??}. 
%%For real-world testing, Cheetah naturally replaced instances of the baseline cloud (Section~\ref{subsec:cloudBL}) with actual cloud applications. %We are now working on systematically evaluating the usability of our simulator in partnership with Cheetah and based on the existing guidelines in the literature~\cite{Rodrigues:17}.

\iffalse
We presented the \lang\ language in a number of meetings to Cheetah's engineers and updated the language based on their feedback. The engineers have shown great interests in using the language to modify the simulator on their own to be able to adapt it to the different IoT product use cases that they test. We are working on systematically evaluating the usability of our simulator in the context of our partnership with Cheetah Networks and based on the literature guidelines~\cite{Rodrigues:17}. 
\fi

\subsection{Lessons Learned}

\mbox{}\indent\textbf{UML profiles versus Xtext.} In most of our past research with industry, we have relied on UML profiles~\cite{UMLProfiles} for addressing domain-specific needs. Examples include a profile for managing safety-certification documents~\cite{Rajwinder:13} (collaboration with the maritime and energy sector), a profile for simulating tax policies~\cite{Soltana:18} (collaboration with the public-service sector), a profile for traceability analysis between requirements and design~\cite{NejatiFSE16} (collaboration with the automotive industry), and a profile for hardware-in-the-loop (HiL) testing of cyber-physical systems~\cite{Shin:18} (collaboration with the satellite industry).

In our current work, we considered UML profiles as an alternative to Xtext. Nonetheless, in view of the contextual factors that we highlight below, we favoured Xtext. First, while the engineers at our partner were familiar with UML, they were not actively using it for product development. We deliberated over a profile-based solution that would mostly hide UML. However, the issue of DSL maintenance lingered. Specifically, the issue is as follows: if the partner company were to independently modify the DSL, considering that UML was not being used, would it be easier for the company to modify an Xtext-based implementation or a profile-based one? Given the context, we deemed an Xtext-based implementation more likely to be maintainable. Another important factor that influenced our decision was whether using UML would lead to a significantly simpler solution. In many situations, tailoring the existing modelling languages in UML is superior to building from scratch and potentially ``reinventing the wheel''. In the context of our current work though, as the conceptualization and algorithms of Section~\ref{subsec:metamodel} emerged, we determined that a lightweight, UML-independent DSL would be sufficient for meeting our analytical goals. GMF~\cite{GMP} was also considered as an alternative, but a textual notation was found to be more palatable to our partner for the analytical goal at hand.

The lesson learned here is that contextual factors, like in many other aspects of software engineering~\cite{Basili:18}, play an important role in choosing a suitable DSL technology that is fit for purpose.

\textbf{Making simulation cost-aware.} We did not initially anticipate simulation costs to be a factor in our DSL grammar design. The ability to constrain memory and CPU usage for containers and virtual machines was provisioned into \lang\ later and in response to our partner's feedback about simulation costs. In particular, our partner noted that, in large IoT simulation and testing campaigns involving hundreds of thousands of simulated devices, solution providers often use externally hosted resources for a fee; the fee is determined based on the CPU and memory limits set for containers and virtual machines. We believe that accounting for cost factors -- in our case, hosting costs -- is important for simulation DSLs, as doing so provides a way to \emph{optimize} the cost of simulation. For example, we are now considering how, through optimization techniques such as search, we can maximize the number of IoT and edge devices that can be simulated while minimizing hosting costs.
%\textbf{Using Simulator for Network Testing.}

%If we connect computers through a public network or via Wifi, we observe significant network loss. This  suggests that our setting could be used for network testing as well although we do not pursue this direction in this paper. This is a use case that our partner is also interested in. 

%Rodrigues et al.~\cite{Rodrigues:17} focuses on evaluation of usability of DSLs by presenting a Systematic Literature Review and proposes a new framework called Usa-DSL. In~\cite{Rodrigues:18} they describes this new framework which contains four phases: Planning, Execution, Analysis and Results. Each phase can be split in several steps including evaluators profile, ethical and legal responsibilities, data types, empirical study methods, evaluation method, metrics, gathering instruments, evaluation instructions, evaluation conduction, data packaging and evaluation reporting. Facing the challenge of few experts in this specialised domain~\cite{Borum:21}, we tailored this framework as not all steps have to be performed. Now we gathered some feedback from a pilot test, and are following Usa-DSL for a  usability evaluation through user observation using interviews and questionnaires.

%When you are doing a systematic test at this level, you need to rule out unwanted behaviours from the simulator before you can actually use the simulator. For that you need proper knobs of the simulator and a systematic approach to tune your simulator. The simulator should not collapse under its own weight. 

\vspace*{-.25em}
\section{Related Work}
\label{sec:related}
We compare our approach with related work in the areas of IoT simulation, IoT testing and domain-specific languages for IoT. 

\textbf{IoT Simulation.} Numerous simulators exist for different components of IoT systems, e.g., IoT sensors and actuators~\cite{Pflanzner:16,NuvIoT}, IoT edge devices~\cite{Jha:20,Truong:21}, IoT networks~\cite{OMNeT++,Cooja,NS-3}, and IoT cloud infrastructure~\cite{Calheiros:2009,Garg:11,Zeng:16}.
Among these, the closest ones to our work are IoT edge simulators. IoTCloudSamples~\cite{Truong:21} 
provides a simluation framework aimed at facilitating the development and operation of edge software systems. IoTSim-Edge~\cite{Jha:20} is an IoT device-to-edge simulator enabling the specification of various IoT-device characteristics, e.g., network connectivity, mobility and energy consumption, and simulating the interactions of IoT devices with edge devices.

The above simulators are focused on covering various features and metrics in the IoT domain. In contrast, \lang\ is not built for feature coverage, but rather to provide a lean, purpose-built simulation framework for cloud scale testing that can simulate, with minimal resources, thousands of devices frequently communicating with the cloud. IoTECS’ symbolic execution of IoT devices at the edge layer and its simulation architecture management are novel and specifically aimed at minimizing simulation-resource needs.

%The above simulators focus primarily on the information  flow at the boundary between IoT devices and the edge.  In contrast, our work is concerned with the information flow between the edge and the cloud layers. Furthermore, the analytical goal we pursue, namely stress testing, is different from the use cases and application scenarios that the previously developed edge simulators consider.  

\textbf{IoT Testing.} Testing IoT systems is a challenging task due to numerous factors, e.g., the sheer scale of IoT systems, the heterogeneity of IoT sensors and actuators, the diversity of network protocols and network topologies in IoT systems, and the tight integration of IoT systems with their environment~\cite{Beilharz:2021}.

Motivated by testing IoT systems, Moawad et al.~\cite{Moawad:15} propose a conceptualization of IoT data at run-time. They then use time-series compression and polynomial segmentation algorithms
to synthesize realistic sensor data based on their proposed conceptualization.

EMU-IoT~\cite{Ramprasad:19} is an IoT testbed based on a microservice architecture. This testbed can emulate IoT devices through containers running on virtual machines, create virtualized gateways, and define orchestrators, monitors and load balancers. EMU-IoT has been applied for various testing purposes, e.g., testing the collection of information about resource consumption across an IoT system.
IOTier~\cite{Nikolaidis:21} is a testbed that enables using resource-constrained containers as IoT components and grouping these components into a device tier, gateway tier and cloud tier. Using IOTier, one can simulate dynamic changes in the operating conditions of IoT systems and thereby test device capabilities, network performance and load-balancing strategies. 
Fogify\cite{Symeonides:20} is an edge testbed that focuses on emulating edge topologies and deployment conditions using containerized descriptions. Fogify further provides a mechanism to alter network quality and inject entity and infrastructure downtime at run-time. 
UiTiOt (version 3)~\cite{Ly-Trong:18} provides a testbed for integrating real IoT devices with emulated devices that run as containers on virtual machines to support large-scale experimentation.
Sendorek et al.~\cite{Sendorek:18} propose the concept of a software-defined IoT test environment.  
Their proposed testbed can define virtual environments, virtual sensors and virtual communication devices as software, thereby allowing the combination of virtual and real equipment for testing IoT application and communication protocols.

None of these works support the testing objective that motivates our work, i.e., scale testing of \emph{cloud} applications. Further, although some of these works use virtualization and containerization methods, none perform systematic experiments that compare containers versus virtual machines and native hosts for simulation. Our experimentation not only demonstrates the applicability of IoTECS, but also the inadequacy of non-hierarchical (native) simulation which does not benefit from our proposed architecture.

\textbf{Domain-Specific Languages (DSLs) for IoT.}
Model-driven engineering (MDE) is a well-established thrust in IoT~\cite{Morin:17,Song:20,Moawad:15}.
DSLs are particularly common as a way to abstract away from the complexity of IoT systems and thereby simplify the construction and analysis of this type of systems.

Sneps{-}Sneppe and Namiot~\cite{Sneppe:15} propose a web-based DSL to simplify data access in IoT web applications. This DSL supports both synchronous and asynchronous data updates as well as communication between processes and sensors.
Tichy et al.~\cite{Tichy:20} present a DSL that supports the declarative specification and execution of IoT system components for quality assurance purposes.
Gomes~et~al.~\cite{Gomes:17} propose EL4IoT, a DSL that aims to simplify the development of IoT-device applications by refactoring the network stack.
Amrani et al.~\cite{Amrani:17} propose IoTDSL for specifying and assembling usage scenarios of IoT systems in a way that would be understandable for end-users without programming expertise. %{\AA}kesson et al.~\cite{Akesson:19} develop a DSL for IoT service composition. This DSL  supports live programming, drag-and-drop, and sequential, alternative and parallel events. %Finally, Cacciagrano and Culmone~\cite{Cacciagrano:20} propose a DSL, named IRON, for programming smart environments based on Event-Condition-Action (ECA) rules~\cite{Cacciagrano:18}. The end goal here is to provide the ability to intercept specific program anomalies and \hbox{application-specific safety issues.}

These existing DSLs have helped us better shape our approach; nevertheless, none have not been designed for our use case, and cannot, without very substantial resources, generate large-scale, lossless edge-cloud traffic. %Further, none provide a DSL.
\section{Conclusions}
\label{sec:con}
In this paper, we proposed a domain-specific language, named \lang, for building simulators that mimic the operation of a large number of edge devices. The main motivating use case for \lang\ is stress testing of IoT cloud applications and systematically measuring how these applications scale. \lang\ was designed, iteratively refined, and empirically evaluated in the context of a collaborative project with industry. \lang\ has been implemented using Xtext and Xtend, and is publicly available~\cite{IoTECS}.

For future work, we would like to enhance \lang\ with high-level constructs that enable a more flexible configuration of the publish-subscribe messaging model in simulators. In addition, we would like to study optimization techniques to optimally balance the total number of simulated IoT devices against simulation costs.

\iffalse
Future work: We note that our simulator can also be used for other use cases such as testing the behaviour of IoT devices and edge devices. For those use cases, however, the simulator cannot be executed on a single machine and has to be executed on a distributed setup. Investigating such use cases are left for future work.

If we connect computers through a public network or via Wifi, we observe significant network loss. This  suggests that our setting could be used for network testing as well although we do not pursue this direction in this paper. This is a use case that our partner is also interested in. 
\fi

\begin{acks}
We gratefully acknowledge funding from Mitacs Accelerate (IT28142), Cheetah Networks, and NSERC of Canada under the Discovery and Discovery Accelerator programs.
\end{acks}

\bibliographystyle{ACM-Reference-Format}
\balance
\bibliography{paper}


\begin{thebibliography}{53}

%%% the Web address.

\ifx \showCODEN    \undefined \def \showCODEN     #1{\unskip}     \fi
\ifx \showDOI      \undefined \def \showDOI       #1{#1}\fi
\ifx \showISBNx    \undefined \def \showISBNx     #1{\unskip}     \fi
\ifx \showISBNxiii \undefined \def \showISBNxiii  #1{\unskip}     \fi
\ifx \showISSN     \undefined \def \showISSN      #1{\unskip}     \fi
\ifx \showLCCN     \undefined \def \showLCCN      #1{\unskip}     \fi
\ifx \shownote     \undefined \def \shownote      #1{#1}          \fi
\ifx \showarticletitle \undefined \def \showarticletitle #1{#1}   \fi
\ifx \showURL      \undefined \def \showURL       {\relax}        \fi
% invisible to TeX
\providecommand\bibfield[2]{#2}
\providecommand\bibinfo[2]{#2}
\providecommand\natexlab[1]{#1}
\providecommand\showeprint[2][]{arXiv:#2}

\bibitem[GMP({[n.\,d.]})]%
        {GMP}
 \bibinfo{year}{[n.\,d.]}\natexlab{}.
\newblock \bibinfo{title}{Graphical Modeling Project (GMP)}.
\newblock \bibinfo{howpublished}{\url{https://www.eclipse.org/modeling/gmp/}}.
\newblock


\bibitem[Coo(2016)]%
        {Cooja}
 \bibinfo{year}{2016}\natexlab{}.
\newblock \bibinfo{title}{Cooja Simulator}.
\newblock
  \bibinfo{howpublished}{\url{https://anrg.usc.edu/contiki/index.php/Cooja_Simulator}}.
\newblock


\bibitem[Ama(2021)]%
        {AmazonIoT}
 \bibinfo{year}{2021}\natexlab{}.
\newblock \bibinfo{title}{Amazon Web Services. IoT Core.}
\newblock
  \bibinfo{howpublished}{\url{https://aws.amazon.com/solutions/implementations/iot-device-simulator/}}.
\newblock


\bibitem[Min(2021)]%
        {Mininet}
 \bibinfo{year}{2021}\natexlab{}.
\newblock \bibinfo{title}{Mininet}.
\newblock \bibinfo{howpublished}{\url{https://mininet.org}}.
\newblock


\bibitem[NS-(2021)]%
        {NS-3}
 \bibinfo{year}{2021}\natexlab{}.
\newblock \bibinfo{title}{NS-3.35}.
\newblock \bibinfo{howpublished}{\url{https://www.nsnam.org}}.
\newblock


\bibitem[IoT(2022)]%
        {IoTECS}
 \bibinfo{year}{2022}\natexlab{}.
\newblock \bibinfo{title}{IoTECS}.
\newblock \bibinfo{howpublished}{\url{https://github.com/JiaLi123456/IoTECS}}.
\newblock


\bibitem[OMN(2022)]%
        {OMNeT++}
 \bibinfo{year}{2022}\natexlab{}.
\newblock \bibinfo{title}{OMNeT++}.
\newblock \bibinfo{howpublished}{\url{https://omnetpp.org/intro/}}.
\newblock


\bibitem[Ahlgren et~al\mbox{.}(2021)]%
        {Ahlgren:21}
\bibfield{author}{\bibinfo{person}{John Ahlgren}, \bibinfo{person}{Kinga
  Bojarczuk}, \bibinfo{person}{Sophia Drossopoulou}, \bibinfo{person}{Inna
  Dvortsova}, \bibinfo{person}{Johann George}, \bibinfo{person}{Natalija
  Gucevska}, \bibinfo{person}{Mark Harman}, \bibinfo{person}{Maria Lomeli},
  \bibinfo{person}{Simon M.~M. Lucas}, \bibinfo{person}{Erik Meijer},
  \bibinfo{person}{Steve Omohundro}, \bibinfo{person}{Rubmary Rojas},
  \bibinfo{person}{Silvia Sapora}, {and} \bibinfo{person}{Norm Zhou}.}
  \bibinfo{year}{2021}\natexlab{}.
\newblock \showarticletitle{Facebook's Cyber-Cyber and Cyber-Physical Digital
  Twins}. In \bibinfo{booktitle}{\emph{{EASE} 2021: Evaluation and Assessment
  in Software Engineering, Trondheim, Norway, June 21-24, 2021}},
  \bibfield{editor}{\bibinfo{person}{Ruzanna Chitchyan},
  \bibinfo{person}{Jingyue Li}, \bibinfo{person}{Barbara Weber}, {and}
  \bibinfo{person}{Tao Yue}} (Eds.). \bibinfo{pages}{1--9}.
\newblock
\urldef\tempurl%
\url{https://doi.org/10.1145/3463274.3463275}
\showDOI{\tempurl}


\bibitem[Al{-}Sultan et~al\mbox{.}(2014)]%
        {Al-Sultan:14}
\bibfield{author}{\bibinfo{person}{Saif~Jamal Al{-}Sultan},
  \bibinfo{person}{Moath~M. Al{-}Doori}, \bibinfo{person}{Ali~H. Al{-}Bayatti},
  {and} \bibinfo{person}{Hussein Zedan}.} \bibinfo{year}{2014}\natexlab{}.
\newblock \showarticletitle{A comprehensive survey on vehicular Ad Hoc
  network}.
\newblock \bibinfo{journal}{\emph{Journal of Network and Computer
  Applications}}  \bibinfo{volume}{37} (\bibinfo{year}{2014}),
  \bibinfo{pages}{380--392}.
\newblock
\urldef\tempurl%
\url{https://doi.org/10.1016/j.jnca.2013.02.036}
\showDOI{\tempurl}


\bibitem[Amrani et~al\mbox{.}(2017)]%
        {Amrani:17}
\bibfield{author}{\bibinfo{person}{Moussa Amrani}, \bibinfo{person}{Fabian
  Gilson}, \bibinfo{person}{Abdelmounaim Debieche}, {and}
  \bibinfo{person}{Vincent Englebert}.} \bibinfo{year}{2017}\natexlab{}.
\newblock \showarticletitle{Towards User-centric DSLs to Manage IoT Systems}.
  In \bibinfo{booktitle}{\emph{Proceedings of the 5th International Conference
  on Model-Driven Engineering and Software Development, {MODELSWARD} 2017,
  Porto, Portugal, February 19-21, 2017}},
  \bibfield{editor}{\bibinfo{person}{Lu{\'{\i}}s~Ferreira Pires},
  \bibinfo{person}{Slimane Hammoudi}, {and} \bibinfo{person}{Bran Selic}}
  (Eds.). \bibinfo{pages}{569--576}.
\newblock
\urldef\tempurl%
\url{https://doi.org/10.5220/0006285405690576}
\showDOI{\tempurl}


\bibitem[Basili et~al\mbox{.}(2018)]%
        {Basili:18}
\bibfield{author}{\bibinfo{person}{Victor~R. Basili},
  \bibinfo{person}{Lionel~C. Briand}, \bibinfo{person}{Domenico Bianculli},
  \bibinfo{person}{Shiva Nejati}, \bibinfo{person}{Fabrizio Pastore}, {and}
  \bibinfo{person}{Mehrdad Sabetzadeh}.} \bibinfo{year}{2018}\natexlab{}.
\newblock \showarticletitle{Software Engineering Research and Industry: {A}
  Symbiotic Relationship to Foster Impact}.
\newblock \bibinfo{journal}{\emph{{IEEE} Software}} \bibinfo{volume}{35},
  \bibinfo{number}{5} (\bibinfo{year}{2018}), \bibinfo{pages}{44--49}.
\newblock
\urldef\tempurl%
\url{https://doi.org/10.1109/MS.2018.290110216}
\showDOI{\tempurl}


\bibitem[Beilharz et~al\mbox{.}(2021)]%
        {Beilharz:2021}
\bibfield{author}{\bibinfo{person}{Jossekin Beilharz}, \bibinfo{person}{Philipp
  Wiesner}, \bibinfo{person}{Arne Boockmeyer}, \bibinfo{person}{Lukas Pirl},
  \bibinfo{person}{Dirk Friedenberger}, \bibinfo{person}{Florian Brokhausen},
  \bibinfo{person}{Ilja Behnke}, \bibinfo{person}{Andreas Polze}, {and}
  \bibinfo{person}{Lauritz Thamsen}.} \bibinfo{year}{2021}\natexlab{}.
\newblock \showarticletitle{Continuously Testing Distributed IoT Systems: An
  Overview of the State of the Art}.
\newblock \bibinfo{journal}{\emph{CoRR}}  \bibinfo{volume}{abs/2112.09580}
  (\bibinfo{year}{2021}).
\newblock


\bibitem[Borg et~al\mbox{.}(2021)]%
        {Borg:21}
\bibfield{author}{\bibinfo{person}{Markus Borg}, \bibinfo{person}{Raja~Ben
  Abdessalem}, \bibinfo{person}{Shiva Nejati},
  \bibinfo{person}{Fran{\c{c}}ois{-}Xavier Jegeden}, {and}
  \bibinfo{person}{Donghwan Shin}.} \bibinfo{year}{2021}\natexlab{}.
\newblock \showarticletitle{Digital Twins Are Not Monozygotic -
  Cross-Replicating {ADAS} Testing in Two Industry-Grade Automotive
  Simulators}. In \bibinfo{booktitle}{\emph{14th {IEEE} Conference on Software
  Testing, Verification and Validation, {ICST} 2021, Porto de Galinhas, Brazil,
  April 12-16, 2021}}. \bibinfo{pages}{383--393}.
\newblock
\urldef\tempurl%
\url{https://doi.org/10.1109/ICST49551.2021.00050}
\showDOI{\tempurl}


\bibitem[Boulis(2011)]%
        {boulis:11}
\bibfield{author}{\bibinfo{person}{Athanassios Boulis}.}
  \bibinfo{year}{2011}\natexlab{}.
\newblock \showarticletitle{Castalia: A simulator for wireless sensor networks
  and body area networks}.
\newblock \bibinfo{journal}{\emph{NICTA: National ICT Australia}}
  \bibinfo{volume}{83} (\bibinfo{year}{2011}), \bibinfo{pages}{7}.
\newblock


\bibitem[Calheiros et~al\mbox{.}(2009)]%
        {Calheiros:2009}
\bibfield{author}{\bibinfo{person}{Rodrigo~N. Calheiros},
  \bibinfo{person}{Rajiv Ranjan}, \bibinfo{person}{C{\'{e}}sar A. F.~De Rose},
  {and} \bibinfo{person}{Rajkumar Buyya}.} \bibinfo{year}{2009}\natexlab{}.
\newblock \showarticletitle{CloudSim: {A} Novel Framework for Modeling and
  Simulation of Cloud Computing Infrastructures and Services}.
\newblock \bibinfo{journal}{\emph{CoRR}}  \bibinfo{volume}{abs/0903.2525}
  (\bibinfo{year}{2009}).
\newblock


\bibitem[Chan(2004)]%
        {Chan:04}
\bibfield{author}{\bibinfo{person}{H.A. Chan}.}
  \bibinfo{year}{2004}\natexlab{}.
\newblock \showarticletitle{Accelerated stress testing for both hardware and
  software}. In \bibinfo{booktitle}{\emph{Annual Symposium Reliability and
  Maintainability, 2004 - RAMS}}. \bibinfo{pages}{346--351}.
\newblock
\urldef\tempurl%
\url{https://doi.org/10.1109/RAMS.2004.1285473}
\showDOI{\tempurl}


\bibitem[{Combs et al.}(2021)]%
        {Wireshark}
\bibfield{author}{\bibinfo{person}{Gerald {Combs et al.}}}
  \bibinfo{year}{2021}\natexlab{}.
\newblock \bibinfo{title}{Wireshark v3.6.0}.
\newblock \bibinfo{howpublished}{\url{https://www.wireshark.org/}}.
\newblock


\bibitem[{Eclipse Foundation, Inc.}(2021a)]%
        {Xtend}
\bibfield{author}{\bibinfo{person}{{Eclipse Foundation, Inc.}}}
  \bibinfo{year}{2021}\natexlab{a}.
\newblock \bibinfo{title}{Xtend v2.25.0}.
\newblock \bibinfo{howpublished}{\url{https://www.eclipse.org/Xtend/}}.
\newblock


\bibitem[{Eclipse Foundation, Inc.}(2021b)]%
        {Xtext}
\bibfield{author}{\bibinfo{person}{{Eclipse Foundation, Inc.}}}
  \bibinfo{year}{2021}\natexlab{b}.
\newblock \bibinfo{title}{Xtext v2.25.0}.
\newblock \bibinfo{howpublished}{\url{https://www.eclipse.org/Xtext/}}.
\newblock


\bibitem[Elazhary(2019)]%
        {Elazhary:19}
\bibfield{author}{\bibinfo{person}{Hanan Elazhary}.}
  \bibinfo{year}{2019}\natexlab{}.
\newblock \showarticletitle{Internet of Things (IoT), mobile cloud, cloudlet,
  mobile IoT, IoT cloud, fog, mobile edge, and edge emerging computing
  paradigms: Disambiguation and research directions}.
\newblock \bibinfo{journal}{\emph{Journal of Network and Computer
  Applications}}  \bibinfo{volume}{128} (\bibinfo{year}{2019}),
  \bibinfo{pages}{105--140}.
\newblock
\urldef\tempurl%
\url{https://doi.org/10.1016/j.jnca.2018.10.021}
\showDOI{\tempurl}


\bibitem[Garg and Buyya(2011)]%
        {Garg:11}
\bibfield{author}{\bibinfo{person}{Saurabh~Kumar Garg} {and}
  \bibinfo{person}{Rajkumar Buyya}.} \bibinfo{year}{2011}\natexlab{}.
\newblock \showarticletitle{NetworkCloudSim: Modelling Parallel Applications in
  Cloud Simulations}. In \bibinfo{booktitle}{\emph{{IEEE} 4th International
  Conference on Utility and Cloud Computing, {UCC} 2011, Melbourne, Australia,
  December 5-8, 2011}}. \bibinfo{pages}{105--113}.
\newblock
\urldef\tempurl%
\url{https://doi.org/10.1109/UCC.2011.24}
\showDOI{\tempurl}


\bibitem[Gomes et~al\mbox{.}(2017)]%
        {Gomes:17}
\bibfield{author}{\bibinfo{person}{Tiago Gomes}, \bibinfo{person}{P. Lopes},
  \bibinfo{person}{J. Alves}, \bibinfo{person}{Pedro Mestre},
  \bibinfo{person}{Jorge Cabral}, \bibinfo{person}{Jo{\~{a}}o~L. Monteiro},
  {and} \bibinfo{person}{Adriano Tavares}.} \bibinfo{year}{2017}\natexlab{}.
\newblock \showarticletitle{A modeling domain-specific language for IoT-enabled
  operating systems}. In \bibinfo{booktitle}{\emph{{IECON} 2017 - 43rd Annual
  Conference of the {IEEE} Industrial Electronics Society, Beijing, China,
  October 29 - November 1, 2017}}. \bibinfo{pages}{3945--3950}.
\newblock
\urldef\tempurl%
\url{https://doi.org/10.1109/IECON.2017.8216675}
\showDOI{\tempurl}


\bibitem[Halegoua(2020)]%
        {Halegoua:20}
\bibfield{author}{\bibinfo{person}{Germaine~R. Halegoua}.}
  \bibinfo{year}{2020}\natexlab{}.
\newblock \bibinfo{booktitle}{\emph{Smart cities}}.
\newblock \bibinfo{publisher}{The MIT Press}, \bibinfo{address}{Cambridge}.
\newblock


\bibitem[Hung(2017)]%
        {IoT17}
\bibfield{author}{\bibinfo{person}{Mark Hung}.}
  \bibinfo{year}{2017}\natexlab{}.
\newblock \bibinfo{title}{Leading the IoT}.
\newblock \bibinfo{howpublished}{Documentation at
  \url{https://www.gartner.com/imagesrv/books/iot/iotEbook_digital.pdf}}.
\newblock


\bibitem[Jha et~al\mbox{.}(2020)]%
        {Jha:20}
\bibfield{author}{\bibinfo{person}{Devki~Nandan Jha}, \bibinfo{person}{Khaled
  Alwasel}, \bibinfo{person}{Areeb Alshoshan}, \bibinfo{person}{Xianghua
  Huang}, \bibinfo{person}{Ranesh~Kumar Naha}, \bibinfo{person}{Sudheer~Kumar
  Battula}, \bibinfo{person}{Saurabh Garg}, \bibinfo{person}{Deepak Puthal},
  \bibinfo{person}{Philip James}, \bibinfo{person}{Albert~Y. Zomaya},
  \bibinfo{person}{Schahram Dustdar}, {and} \bibinfo{person}{Rajiv Ranjan}.}
  \bibinfo{year}{2020}\natexlab{}.
\newblock \showarticletitle{IoTSim-Edge: {A} simulation framework for modeling
  the behavior of Internet of Things and edge computing environments}.
\newblock \bibinfo{journal}{\emph{Software - Practice and Experience}}
  \bibinfo{volume}{50} (\bibinfo{year}{2020}), \bibinfo{pages}{844--867}.
\newblock
\urldef\tempurl%
\url{https://doi.org/10.1002/spe.2787}
\showDOI{\tempurl}


\bibitem[Kert{\'{e}}sz et~al\mbox{.}(2019)]%
        {Kertesz:19}
\bibfield{author}{\bibinfo{person}{Attila Kert{\'{e}}sz},
  \bibinfo{person}{Tamas Pflanzner}, {and} \bibinfo{person}{Tibor
  Gyim{\'{o}}thy}.} \bibinfo{year}{2019}\natexlab{}.
\newblock \showarticletitle{A Mobile IoT Device Simulator for IoT-Fog-Cloud
  Systems}.
\newblock \bibinfo{journal}{\emph{Journal of Grid Computing}}
  \bibinfo{volume}{17}, \bibinfo{number}{3} (\bibinfo{year}{2019}),
  \bibinfo{pages}{529--551}.
\newblock
\urldef\tempurl%
\url{https://doi.org/10.1007/s10723-018-9468-9}
\showDOI{\tempurl}


\bibitem[Kliazovich et~al\mbox{.}(2012)]%
        {Kliazovich:12}
\bibfield{author}{\bibinfo{person}{Dzmitry Kliazovich}, \bibinfo{person}{Pascal
  Bouvry}, {and} \bibinfo{person}{Samee~Ullah Khan}.}
  \bibinfo{year}{2012}\natexlab{}.
\newblock \showarticletitle{GreenCloud: a packet-level simulator of
  energy-aware cloud computing data centers}.
\newblock \bibinfo{journal}{\emph{The Journal of Supercomputing}}
  \bibinfo{volume}{62}, \bibinfo{number}{3} (\bibinfo{year}{2012}),
  \bibinfo{pages}{1263--1283}.
\newblock
\urldef\tempurl%
\url{https://doi.org/10.1007/s11227-010-0504-1}
\showDOI{\tempurl}


\bibitem[Li et~al\mbox{.}(2022)]%
        {Li:22}
\bibfield{author}{\bibinfo{person}{Jia Li}, \bibinfo{person}{Shiva Nejati},
  {and} \bibinfo{person}{Mehrdad Sabetzadeh}.} \bibinfo{year}{2022}\natexlab{}.
\newblock \showarticletitle{Learning Self-adaptations for IoT Networks: A
  Genetic Programming Approach}. In \bibinfo{booktitle}{\emph{{SEAMS} '22:
  {IEEE/ACM} 17th International Symposium on Software Engineering for Adaptive
  and Self-Managing Systems, Pittsburgh, PA, USA, May 21 - 29, 2022}}.
\newblock


\bibitem[Logistics(2019)]%
        {NuvIoT}
\bibfield{author}{\bibinfo{person}{Software Logistics}.}
  \bibinfo{year}{2019}\natexlab{}.
\newblock \bibinfo{title}{NuvIoT IoT Simulator}.
\newblock \bibinfo{howpublished}{\url{https://www.nuviot.com}}.
\newblock


\bibitem[Ly{-}Trong et~al\mbox{.}(2018)]%
        {Ly-Trong:18}
\bibfield{author}{\bibinfo{person}{Nhan Ly{-}Trong}, \bibinfo{person}{Chuong
  Dang{-}Le{-}Bao}, \bibinfo{person}{Dang Huynh{-}Van}, {and}
  \bibinfo{person}{Quan~Le Trung}.} \bibinfo{year}{2018}\natexlab{}.
\newblock \showarticletitle{UiTiOt v3: {A} Hybrid Testbed for Evaluation of
  Large-Scale IoT Networks}. In \bibinfo{booktitle}{\emph{Proceedings of the
  Ninth International Symposium on Information and Communication Technology,
  SoICT 2018, Danang City, Vietnam, December 06-07, 2018}}.
  \bibinfo{pages}{155--162}.
\newblock
\urldef\tempurl%
\url{https://doi.org/10.1145/3287921.3287935}
\showDOI{\tempurl}


\bibitem[Moawad et~al\mbox{.}(2015)]%
        {Moawad:15}
\bibfield{author}{\bibinfo{person}{Assaad Moawad}, \bibinfo{person}{Thomas
  Hartmann}, \bibinfo{person}{Fran{\c{c}}ois Fouquet},
  \bibinfo{person}{Gr{\'{e}}gory Nain}, \bibinfo{person}{Jacques Klein}, {and}
  \bibinfo{person}{Yves~Le Traon}.} \bibinfo{year}{2015}\natexlab{}.
\newblock \showarticletitle{Beyond discrete modeling: {A} continuous and
  efficient model for IoT}. In \bibinfo{booktitle}{\emph{18th {ACM/IEEE}
  International Conference on Model Driven Engineering Languages and Systems,
  MoDELS 2015, Ottawa, ON, Canada, September 30 - October 2, 2015}},
  \bibfield{editor}{\bibinfo{person}{Timothy Lethbridge},
  \bibinfo{person}{Jordi Cabot}, {and} \bibinfo{person}{Alexander Egyed}}
  (Eds.). \bibinfo{pages}{90--99}.
\newblock
\urldef\tempurl%
\url{https://doi.org/10.1109/MODELS.2015.7338239}
\showDOI{\tempurl}


\bibitem[Morin et~al\mbox{.}(2017)]%
        {Morin:17}
\bibfield{author}{\bibinfo{person}{Brice Morin}, \bibinfo{person}{Nicolas
  Harrand}, {and} \bibinfo{person}{Franck Fleurey}.}
  \bibinfo{year}{2017}\natexlab{}.
\newblock \showarticletitle{Model-Based Software Engineering to Tame the IoT
  Jungle}.
\newblock \bibinfo{journal}{\emph{{IEEE} Software}} \bibinfo{volume}{34},
  \bibinfo{number}{1} (\bibinfo{year}{2017}), \bibinfo{pages}{30--36}.
\newblock
\urldef\tempurl%
\url{https://doi.org/10.1109/MS.2017.11}
\showDOI{\tempurl}


\bibitem[Nejati et~al\mbox{.}(2016)]%
        {NejatiFSE16}
\bibfield{author}{\bibinfo{person}{Shiva Nejati}, \bibinfo{person}{Mehrdad
  Sabetzadeh}, \bibinfo{person}{Chetan Arora}, \bibinfo{person}{Lionel~C.
  Briand}, {and} \bibinfo{person}{Felix Mandoux}.}
  \bibinfo{year}{2016}\natexlab{}.
\newblock \showarticletitle{Automated change impact analysis between SysML
  models of requirements and design}. In \bibinfo{booktitle}{\emph{Proceedings
  of the 24th {ACM} {SIGSOFT} International Symposium on Foundations of
  Software Engineering, {FSE} 2016, Seattle, WA, USA, November 13-18, 2016}},
  \bibfield{editor}{\bibinfo{person}{Thomas Zimmermann}, \bibinfo{person}{Jane
  Cleland{-}Huang}, {and} \bibinfo{person}{Zhendong Su}} (Eds.).
  \bibinfo{publisher}{{ACM}}, \bibinfo{pages}{242--253}.
\newblock
\urldef\tempurl%
\url{https://doi.org/10.1145/2950290.2950293}
\showDOI{\tempurl}


\bibitem[Nikolaidis et~al\mbox{.}(2021)]%
        {Nikolaidis:21}
\bibfield{author}{\bibinfo{person}{Fotis Nikolaidis}, \bibinfo{person}{Manolis
  Marazakis}, {and} \bibinfo{person}{Angelos Bilas}.}
  \bibinfo{year}{2021}\natexlab{}.
\newblock \showarticletitle{IOTier: {A} Virtual Testbed to evaluate systems for
  IoT environments}. In \bibinfo{booktitle}{\emph{21st {IEEE/ACM} International
  Symposium on Cluster, Cloud and Internet Computing, CCGrid 2021, Melbourne,
  Australia, May 10-13, 2021}}, \bibfield{editor}{\bibinfo{person}{Laurent
  Lef{\`{e}}vre}, \bibinfo{person}{Stacy Patterson},
  \bibinfo{person}{Young~Choon Lee}, \bibinfo{person}{Haiying Shen},
  \bibinfo{person}{Shashikant Ilager}, \bibinfo{person}{Mohammad Goudarzi},
  \bibinfo{person}{Adel~Nadjaran Toosi}, {and} \bibinfo{person}{Rajkumar
  Buyya}} (Eds.). \bibinfo{pages}{676--683}.
\newblock
\urldef\tempurl%
\url{https://doi.org/10.1109/CCGrid51090.2021.00081}
\showDOI{\tempurl}


\bibitem[(OMG)(2017)]%
        {UMLProfiles}
\bibfield{author}{\bibinfo{person}{Object Management~Group (OMG)}.}
  \bibinfo{year}{2017}\natexlab{}.
\newblock \bibinfo{title}{Unified Modeling Language (UML) Specification Version
  2.5.1}.
\newblock \bibinfo{howpublished}{\url{https://www.omg.org/spec/UML/2.5.1}}.
\newblock


\bibitem[{\"{O}}sterlind et~al\mbox{.}(2006)]%
        {Osterlind:06}
\bibfield{author}{\bibinfo{person}{Fredrik {\"{O}}sterlind},
  \bibinfo{person}{Adam Dunkels}, \bibinfo{person}{Joakim Eriksson},
  \bibinfo{person}{Niclas Finne}, {and} \bibinfo{person}{Thiemo Voigt}.}
  \bibinfo{year}{2006}\natexlab{}.
\newblock \showarticletitle{Cross-Level Sensor Network Simulation with
  {COOJA}}. In \bibinfo{booktitle}{\emph{{LCN} 2006, The 31st Annual {IEEE}
  Conference on Local Computer Networks, Tampa, Florida, USA, 14-16 November
  2006}}. \bibinfo{pages}{641--648}.
\newblock
\urldef\tempurl%
\url{https://doi.org/10.1109/LCN.2006.322172}
\showDOI{\tempurl}


\bibitem[Panesar{-}Walawege et~al\mbox{.}(2013)]%
        {Rajwinder:13}
\bibfield{author}{\bibinfo{person}{Rajwinder~Kaur Panesar{-}Walawege},
  \bibinfo{person}{Mehrdad Sabetzadeh}, {and} \bibinfo{person}{Lionel~C.
  Briand}.} \bibinfo{year}{2013}\natexlab{}.
\newblock \showarticletitle{Supporting the verification of compliance to safety
  standards via model-driven engineering: Approach, tool-support and empirical
  validation}.
\newblock \bibinfo{journal}{\emph{Information and Software Technology}}
  \bibinfo{volume}{55}, \bibinfo{number}{5} (\bibinfo{year}{2013}),
  \bibinfo{pages}{836--864}.
\newblock
\urldef\tempurl%
\url{https://doi.org/10.1016/j.infsof.2012.11.009}
\showDOI{\tempurl}


\bibitem[Pflanzner et~al\mbox{.}(2016)]%
        {Pflanzner:16}
\bibfield{author}{\bibinfo{person}{Tamas Pflanzner}, \bibinfo{person}{Attila
  Kert{\'{e}}sz}, \bibinfo{person}{Bart Spinnewyn}, {and}
  \bibinfo{person}{Steven Latr{\'{e}}}.} \bibinfo{year}{2016}\natexlab{}.
\newblock \showarticletitle{MobIoTSim: Towards a Mobile IoT Device Simulator}.
  In \bibinfo{booktitle}{\emph{4th {IEEE} International Conference on Future
  Internet of Things and Cloud Workshops, FiCloud Workshops 2016, Vienna,
  Austria, August 22-24, 2016}}, \bibfield{editor}{\bibinfo{person}{Muhammad
  Younas}, \bibinfo{person}{Irfan Awan}, {and} \bibinfo{person}{Joyce~El
  Haddad}} (Eds.). \bibinfo{pages}{21--27}.
\newblock
\urldef\tempurl%
\url{https://doi.org/10.1109/W-FiCloud.2016.21}
\showDOI{\tempurl}


\bibitem[Ramprasad et~al\mbox{.}(2019)]%
        {Ramprasad:19}
\bibfield{author}{\bibinfo{person}{Brian Ramprasad}, \bibinfo{person}{Marios
  Fokaefs}, \bibinfo{person}{Joydeep Mukherjee}, {and} \bibinfo{person}{Marin
  Litoiu}.} \bibinfo{year}{2019}\natexlab{}.
\newblock \showarticletitle{EMU-IoT - {A} Virtual Internet of Things Lab}. In
  \bibinfo{booktitle}{\emph{2019 {IEEE} International Conference on Autonomic
  Computing, {ICAC} 2019, Ume{\aa}, Sweden, June 16-20, 2019}}.
  \bibinfo{pages}{73--83}.
\newblock
\urldef\tempurl%
\url{https://doi.org/10.1109/ICAC.2019.00019}
\showDOI{\tempurl}


\bibitem[Scheffer et~al\mbox{.}(1995)]%
        {Scheffer:95}
\bibfield{author}{\bibinfo{person}{M. Scheffer}, \bibinfo{person}{J.M. Baveco},
  \bibinfo{person}{D.L. DeAngelis}, \bibinfo{person}{K.A. Rose}, {and}
  \bibinfo{person}{E.H. {van Nes}}.} \bibinfo{year}{1995}\natexlab{}.
\newblock \showarticletitle{Super-individuals a simple solution for modelling
  large populations on an individual basis}.
\newblock \bibinfo{journal}{\emph{Ecological Modelling}} \bibinfo{volume}{80},
  \bibinfo{number}{2} (\bibinfo{year}{1995}), \bibinfo{pages}{161--170}.
\newblock
\urldef\tempurl%
\url{https://doi.org/10.1016/0304-3800(94)00055-M}
\showDOI{\tempurl}


\bibitem[Sendorek et~al\mbox{.}(2018)]%
        {Sendorek:18}
\bibfield{author}{\bibinfo{person}{Joanna Sendorek}, \bibinfo{person}{Tomasz
  Szydlo}, {and} \bibinfo{person}{Robert Brzoza{-}Woch}.}
  \bibinfo{year}{2018}\natexlab{}.
\newblock \showarticletitle{Software-Defined Virtual Testbed for IoT Systems}.
\newblock \bibinfo{journal}{\emph{Wireless Communications and Mobile
  Computing}}  \bibinfo{volume}{2018} (\bibinfo{year}{2018}),
  \bibinfo{pages}{1068261:1--1068261:11}.
\newblock
\urldef\tempurl%
\url{https://doi.org/10.1155/2018/1068261}
\showDOI{\tempurl}


\bibitem[Shin et~al\mbox{.}(2018)]%
        {Shin:18}
\bibfield{author}{\bibinfo{person}{Seung~Yeob Shin}, \bibinfo{person}{Karim
  Chaouch}, \bibinfo{person}{Shiva Nejati}, \bibinfo{person}{Mehrdad
  Sabetzadeh}, \bibinfo{person}{Lionel~C. Briand}, {and} \bibinfo{person}{Frank
  Zimmer}.} \bibinfo{year}{2018}\natexlab{}.
\newblock \showarticletitle{{HITECS:} {A} {UML} Profile and Analysis Framework
  for Hardware-in-the-Loop Testing of Cyber Physical Systems}. In
  \bibinfo{booktitle}{\emph{Proceedings of the 21th {ACM/IEEE} International
  Conference on Model Driven Engineering Languages and Systems, {MODELS} 2018,
  Copenhagen, Denmark, October 14-19, 2018}},
  \bibfield{editor}{\bibinfo{person}{Andrzej Wasowski},
  \bibinfo{person}{Richard~F. Paige}, {and} \bibinfo{person}{{\O}ystein
  Haugen}} (Eds.). \bibinfo{publisher}{{ACM}}, \bibinfo{pages}{357--367}.
\newblock
\urldef\tempurl%
\url{https://doi.org/10.1145/3239372.3239382}
\showDOI{\tempurl}


\bibitem[Shin et~al\mbox{.}(2020)]%
        {Shin:20}
\bibfield{author}{\bibinfo{person}{Seung~Yeob Shin}, \bibinfo{person}{Shiva
  Nejati}, \bibinfo{person}{Mehrdad Sabetzadeh}, \bibinfo{person}{Lionel~C.
  Briand}, \bibinfo{person}{Chetan Arora}, {and} \bibinfo{person}{Frank
  Zimmer}.} \bibinfo{year}{2020}\natexlab{}.
\newblock \showarticletitle{Dynamic adaptation of software-defined networks for
  IoT systems: a search-based approach}. In \bibinfo{booktitle}{\emph{{SEAMS}
  '20: {IEEE/ACM} 15th International Symposium on Software Engineering for
  Adaptive and Self-Managing Systems, Seoul, Republic of Korea, 29 June - 3
  July, 2020}}, \bibfield{editor}{\bibinfo{person}{Shinichi Honiden},
  \bibinfo{person}{Elisabetta~Di Nitto}, {and} \bibinfo{person}{Radu
  Calinescu}} (Eds.). \bibinfo{pages}{137--148}.
\newblock
\urldef\tempurl%
\url{https://doi.org/10.1145/3387939.3391603}
\showDOI{\tempurl}


\bibitem[Sneps{-}Sneppe and Namiot(2015)]%
        {Sneppe:15}
\bibfield{author}{\bibinfo{person}{Manfred Sneps{-}Sneppe} {and}
  \bibinfo{person}{Dmitry Namiot}.} \bibinfo{year}{2015}\natexlab{}.
\newblock \showarticletitle{On Web-based Domain-Specific Language for Internet
  of Things}.
\newblock \bibinfo{journal}{\emph{CoRR}}  \bibinfo{volume}{abs/1505.06713}
  (\bibinfo{year}{2015}).
\newblock


\bibitem[Soltana et~al\mbox{.}(2018)]%
        {Soltana:18}
\bibfield{author}{\bibinfo{person}{Ghanem Soltana}, \bibinfo{person}{Nicolas
  Sannier}, \bibinfo{person}{Mehrdad Sabetzadeh}, {and}
  \bibinfo{person}{Lionel~C. Briand}.} \bibinfo{year}{2018}\natexlab{}.
\newblock \showarticletitle{Model-based simulation of legal policies:
  framework, tool support, and validation}.
\newblock \bibinfo{journal}{\emph{Software and Systems Modeling}}
  \bibinfo{volume}{17}, \bibinfo{number}{3} (\bibinfo{year}{2018}),
  \bibinfo{pages}{851--883}.
\newblock
\urldef\tempurl%
\url{https://doi.org/10.1007/s10270-016-0542-0}
\showDOI{\tempurl}


\bibitem[Song et~al\mbox{.}(2020)]%
        {Song:20}
\bibfield{author}{\bibinfo{person}{Hui Song}, \bibinfo{person}{Rustem Dautov},
  \bibinfo{person}{Nicolas Ferry}, \bibinfo{person}{Arnor Solberg}, {and}
  \bibinfo{person}{Franck Fleurey}.} \bibinfo{year}{2020}\natexlab{}.
\newblock \showarticletitle{Model-based fleet deployment of edge computing
  applications}. In \bibinfo{booktitle}{\emph{MoDELS '20: {ACM/IEEE} 23rd
  International Conference on Model Driven Engineering Languages and Systems,
  Virtual Event, Canada, 18-23 October, 2020}},
  \bibfield{editor}{\bibinfo{person}{Eugene Syriani},
  \bibinfo{person}{Houari~A. Sahraoui}, \bibinfo{person}{Juan de~Lara}, {and}
  \bibinfo{person}{Silvia Abrah{\~{a}}o}} (Eds.). \bibinfo{pages}{132--142}.
\newblock
\urldef\tempurl%
\url{https://doi.org/10.1145/3365438.3410951}
\showDOI{\tempurl}


\bibitem[Sonmez et~al\mbox{.}(2017)]%
        {Sonmez:17}
\bibfield{author}{\bibinfo{person}{Cagatay Sonmez}, \bibinfo{person}{Atay
  Ozgovde}, {and} \bibinfo{person}{Cem Ersoy}.}
  \bibinfo{year}{2017}\natexlab{}.
\newblock \showarticletitle{EdgeCloudSim: An environment for performance
  evaluation of Edge Computing systems}. In \bibinfo{booktitle}{\emph{Second
  International Conference on Fog and Mobile Edge Computing, {FMEC} 2017,
  Valencia, Spain, May 8-11, 2017}}. \bibinfo{pages}{39--44}.
\newblock
\urldef\tempurl%
\url{https://doi.org/10.1109/FMEC.2017.7946405}
\showDOI{\tempurl}


\bibitem[Symeonides et~al\mbox{.}(2020)]%
        {Symeonides:20}
\bibfield{author}{\bibinfo{person}{Moysis Symeonides},
  \bibinfo{person}{Zacharias Georgiou}, \bibinfo{person}{Demetris Trihinas},
  \bibinfo{person}{George Pallis}, {and} \bibinfo{person}{Marios~D.
  Dikaiakos}.} \bibinfo{year}{2020}\natexlab{}.
\newblock \showarticletitle{Fogify: {A} Fog Computing Emulation Framework}. In
  \bibinfo{booktitle}{\emph{5th {IEEE/ACM} Symposium on Edge Computing, {SEC}
  2020, San Jose, CA, USA, November 12-14, 2020}}. \bibinfo{pages}{42--54}.
\newblock
\urldef\tempurl%
\url{https://doi.org/10.1109/SEC50012.2020.00011}
\showDOI{\tempurl}


\bibitem[Tichy et~al\mbox{.}(2020)]%
        {Tichy:20}
\bibfield{author}{\bibinfo{person}{Matthias Tichy}, \bibinfo{person}{Jakob
  Pietron}, \bibinfo{person}{David M{\"{o}}dinger}, \bibinfo{person}{Katharina
  Juhnke}, {and} \bibinfo{person}{Franz~J. Hauck}.}
  \bibinfo{year}{2020}\natexlab{}.
\newblock \showarticletitle{Experiences with an Internal {DSL} in the IoT
  Domain}. In \bibinfo{booktitle}{\emph{{STAF} 2020 Workshop Proceedings: 4th
  Workshop on Model-Driven Engineering for the Internet-of-Things, 1st
  International Workshop on Modeling Smart Cities, and 5th International
  Workshop on Open and Original Problems in Software Language Engineering
  co-located with Software Technologies: Applications and Foundations
  federation of conferences {(STAF} 2020), Bergen, Norway, June 22-26, 2020}},
  \bibfield{editor}{\bibinfo{person}{Loli Burgue{\~{n}}o} {and}
  \bibinfo{person}{Lars~Michael Kristensen}} (Eds.). \bibinfo{pages}{22--34}.
\newblock


\bibitem[Truong(2021)]%
        {Truong:21}
\bibfield{author}{\bibinfo{person}{Hong~Linh Truong}.}
  \bibinfo{year}{2021}\natexlab{}.
\newblock \showarticletitle{Using IoTCloudSamples as a software framework for
  simulations of edge computing scenarios}.
\newblock \bibinfo{journal}{\emph{Internet Things}}  \bibinfo{volume}{14}
  (\bibinfo{year}{2021}), \bibinfo{pages}{100383}.
\newblock
\urldef\tempurl%
\url{https://doi.org/10.1016/j.iot.2021.100383}
\showDOI{\tempurl}


\bibitem[Wang et~al\mbox{.}(2021)]%
        {Wang:21}
\bibfield{author}{\bibinfo{person}{Xiaoli Wang}, \bibinfo{person}{Bharadwaj
  Veeravalli}, {and} \bibinfo{person}{Jiaming Song}.}
  \bibinfo{year}{2021}\natexlab{}.
\newblock \showarticletitle{Multi-Installment Scheduling for Large-Scale
  Workload Computation with Result Retrieval}.
\newblock \bibinfo{journal}{\emph{Neurocomputing}}  \bibinfo{volume}{458}
  (\bibinfo{year}{2021}), \bibinfo{pages}{579--591}.
\newblock
\urldef\tempurl%
\url{https://doi.org/10.1016/j.neucom.2020.03.124}
\showDOI{\tempurl}


\bibitem[Zeng et~al\mbox{.}(2017)]%
        {Zeng:17}
\bibfield{author}{\bibinfo{person}{Xuezhi Zeng}, \bibinfo{person}{Saurabh~Kumar
  Garg}, \bibinfo{person}{Peter Strazdins}, \bibinfo{person}{Prem~Prakash
  Jayaraman}, \bibinfo{person}{Dimitrios Georgakopoulos}, {and}
  \bibinfo{person}{Rajiv Ranjan}.} \bibinfo{year}{2017}\natexlab{}.
\newblock \showarticletitle{IOTSim: {A} simulator for analysing IoT
  applications}.
\newblock \bibinfo{journal}{\emph{Journal of Systems Architecture: Embedded
  Software Design}}  \bibinfo{volume}{72} (\bibinfo{year}{2017}),
  \bibinfo{pages}{93--107}.
\newblock
\urldef\tempurl%
\url{https://doi.org/10.1016/j.sysarc.2016.06.008}
\showDOI{\tempurl}


\bibitem[Zeng et~al\mbox{.}(2016)]%
        {Zeng:16}
\bibfield{author}{\bibinfo{person}{Xuezhi Zeng}, \bibinfo{person}{Saurabh~Kumar
  Garg}, \bibinfo{person}{Peter~E. Strazdins}, \bibinfo{person}{Prem~Prakash
  Jayaraman}, \bibinfo{person}{Dimitrios Georgakopoulos}, {and}
  \bibinfo{person}{Rajiv Ranjan}.} \bibinfo{year}{2016}\natexlab{}.
\newblock \showarticletitle{IOTSim: a Cloud based Simulator for Analysing IoT
  Applications}.
\newblock \bibinfo{journal}{\emph{CoRR}}  \bibinfo{volume}{abs/1602.06488}
  (\bibinfo{year}{2016}).
\newblock


\end{thebibliography}

%\bibliographystyle{IEEEtran}
%\bibliography{IEEEabrv, paper}
\end{document}